\documentclass[lettersize,journal]{IEEEtran}
\pdfoutput=1
\usepackage{cite}
\usepackage{amsmath,amsfonts}
\usepackage{algorithmic}
\usepackage{algorithm}
\usepackage{array}
\usepackage[caption=false,font=footnotesize,labelfont=rm,textfont=rm,subrefformat=parens]{subfig}
\usepackage{textcomp}
\usepackage{stfloats}
\usepackage{url}
\usepackage{verbatim}
\usepackage{graphicx}
\usepackage{xcolor}
\usepackage{multirow}   
\usepackage[colorlinks, linkcolor=black, anchorcolor=black, citecolor=black]{hyperref}

\begin{document}
	\title{Joint User Scheduling and Precoding for RIS-Aided MU-MISO Systems: A MADRL Approach}
	
	\author{Yangjing Wang, Xiao Li,~\IEEEmembership{Member,~IEEE,} Xinping Yi,~\IEEEmembership{Member,~IEEE,} and Shi Jin,~\IEEEmembership{Fellow,~IEEE}
		\thanks{Y. Wang, X. Li, X. Yi and S. Jin are with the National Mobile Communications Research Laboratory, Southeast University, Nanjing 210096, China (e-mail: {\{yangjingwang, li\_xiao, xyi, jinshi\}@seu.edu.cn}).}
		\thanks{A preliminary version was presented in part at the IEEE ICCT 2023 \cite{wang2023multiagent}.}
	}
	
	
	
	\maketitle
	\begin{abstract}
		 With the increasing demand for spectrum efficiency and energy efficiency, reconfigurable intelligent surfaces (RISs) have attracted massive attention due to its low-cost and capability of controlling wireless environment. 
		 However, there is still a lack of treatments to deal with the growth of the number of users and RIS elements, which may incur performance degradation or 
		 computational complexity explosion.		 
		 In this paper, we investigate the joint optimization of user scheduling and precoding for distributed RIS-aided communication systems. 
		 Firstly, we propose an optimization-based numerical method to obtain suboptimal solutions with the aid of the approximation of ergodic sum rate.
		 Secondly, to reduce the computational complexity caused by the high dimensionality, we propose a data-driven scalable and generalizable multi-agent deep reinforcement learning (MADRL) framework with the aim to maximize the ergodic sum rate approximation through the cooperation of all agents. 
		 Further, we propose a novel dynamic working process exploiting the trained MADRL algorithm, which enables distributed RISs to configure their own passive precoding independently.
		 Simulation results show that our algorithm substantially reduces the computational complexity by a time reduction of three orders of magnitude at the cost of 3\% performance degradation, compared with the optimization-based method, and achieves 6\% performance improvement over the state-of-the-art MADRL algorithms.
	\end{abstract}
	
	
	\section{Introduction}	
	\IEEEPARstart{W}{ith} the evolution of wireless communication systems from the fifth generation (5G) to 6G, the key enabling technologies have become a hotspot of research, including reconfigurable intelligent surfaces (RISs)\cite{wu2020towards}, integrated sensing and communication (ISAC)\cite{liu2022integrated}, terahertz (THz) communications\cite{akyildiz2022terahertz} and so on. Among them, RISs have attracted massive attention in both academia and industry, as an effective means to control wireless propagation environment with low power consumption \cite{wang2022reconfigurable} to improve communication coverage and throughput \cite{sang2022coverage}.
	Specifically, RISs are equipped with a large number of passive elements, each of which can alter the amplitude or phase shift of the incident signal through a smart controller. By coordinating the reflection of all elements, the propagation environment can be flexibly configured, which improves the efficiency and reliability of wireless communication. 
	As an auxiliary device for wireless networks, RISs are highly flexible and easily attached to walls or ceilings. Thus, their applications in wireless communication systems are extensively investigated.
	
	\subsection{Related Works}
	\textbf{Link-level optimization.} To fully harvest the advantages of RISs, link-level optimization for RIS-aided communication systems has been intensively considered. Alternative optimization (AO) methods have been widely used to jointly optimize the active precoding at base station (BS) and the passive precoding at RIS with various objectives, including sum rate maximization \cite{guo2020weighted}, power minimization \cite{wu2019intelligent}, energy efficiency maximization \cite{huang2018reconfigurable} and user fairness maximization \cite{li2023risenhanced}. To overcome the limitations of single RIS-aided communication systems, such as limited coverage area and gain on system performance, distributed RIS-aided systems came into view \cite{zhang2022reconfigurable, li2020weighted, abrardo2021intelligent, gao2021distributed}.
	Firstly, the advantages of distributed RIS deployments over centralized deployments are validated in \cite{zhang2022reconfigurable} exploiting outdated channel state information (CSI).
	Based on instantaneous CSI, an AO method that combines the Lagrangian method and  Riemannian conjugate gradient method was proposed in \cite{li2020weighted}. 
	To reduce the overhead of the phase shift optimization, statistical CSI and instantaneous CSI were utilized to design the passive precoding over large time intervals, and the active precoding over short time intervals in \cite{abrardo2021intelligent}, respectively. 
	To further reduce the overhead and improve the robustness of the system, a low-complexity design for RISs exploiting only statistical CSI was investigated for single-user systems in \cite{gao2021distributed}. 
	
	\textbf{Cross-layer optimization.} In practical overloaded multi-user systems, with the consideration of the system resource limits and data processing capabilities, user scheduling is essential in system design as the number of users increases. Thus, the cross-layer optimization of active precoding at BS, passive precoding at RIS and user scheduling, needs to be considered for RIS-aided communication systems \cite{nadeem2021opportunistic, jiang2023joint, mei2021performance, liu2023joint, amiriara2022irsuser, Al2022Reconfigurable}.
	In \cite{nadeem2021opportunistic}, the proportional fair scheduling and opportunistic beamforming were utilized to design the user scheduling and active precoding for single-input single-output (SISO) systems. 
	For more general multi-cell multi-input single-output (MISO) systems, the design of passive percoding follows the principle of maximizing the effective signal, and then the user pair with the highest correlation coefficient was scheduled in \cite{jiang2023joint}. 
	These heuristic methods have relatively low computational complexity at the expense of performance loss.
	With the fixed BS-user association, the optimal RIS-user association was obtained via the branch-and-bound or successive refinement algorithm in \cite{mei2021performance}. 
	Further, the BS-RIS-user association and beamforming design were decomposed into two subproblems, and solved by the block coordinate descent (BCD) method in \cite{liu2023joint}. 
	While achieving high performance, these numerical optimization methods also lead to extremely high computational complexity. 
	To balance the performance and complexity, artificial intelligence (AI)-based algorithms have been introduced. In \cite{amiriara2022irsuser}, labeled datasets generated by numerical methods were used to train supervised learning algorithms for the cross-layer optimization. 
	In \cite{Al2022Reconfigurable}, deep reinforcement learning (DRL) and BCD methods were employed to optimize the user scheduling and passive precoding, respectively.
	
	\textbf{AI-based optimization.} Thanks to the great advantages of AI in dealing with the nonlinear and non-convex optimization problems, extensive researches on RIS-aided systems based on machine learning, including supervised learning\cite{taha2021enabling, hu2021reconfigurable}, unsupervised learning\cite{song2021unsupervised, jin2024lowcomplexity} and reinforcement learning\cite{feng2020deep, zhang2023a} have been conducted.
	To be specific, a supervised learning algorithm was proposed to learn the direct mapping between sampled CSI and the optimal precoding at RIS in \cite{taha2021enabling}. 
	To reduce the complexity of algorithms, a location-based deep learning architecture was developed, requiring only user locations as input features in \cite{hu2021reconfigurable}.
	Supervised learning utilizes numerical optimization methods to obtain labeled datasets, which greatly increases the complexity of dataset acquisition. 
	In \cite{song2021unsupervised}, data-driven unsupervised learning was utilized to optimize the joint precoding. 
	To further improve the robustness and interpretability of the algorithm, model-driven unsupervised learning was designed and extended to the situation with imperfect CSI in \cite{jin2024lowcomplexity}.
	Moreover, a deep deterministic policy gradient (DDPG) method was employed to optimize the passive precoding exploiting instantaneous CSI in \cite{feng2020deep}. 
	Under two-timescale transmission, the joint precoding was further designed by DRL methods in \cite{zhang2023a}. 
	As the number of RIS elements increases, the curse of dimensionality will cause performance degradation of unsupervised learning and DRL algorithms.
	In comparison, multi-agent DRL (MADRL) algorithms can significantly reduce the impact of dimension on performance and enable the distributed deployment of algorithms.
	In \cite{wang2021hybrid}, a multi-agent DDPG (MADDPG) based hybrid precoding design for mmWave systems was proposed, which achieved better exploration efficiency and convergence speed than single-agent DRL methods. In \cite{naderializadeh2021resource}, a MADRL algorithm was employed to jointly optimize the user scheduling and power control for multi-cell systems in a distributed manner.
	
	\subsection{Motivations and Contributions}
	In practical overloaded multi-user systems, user scheduling becomes essential in system design due to resource constraints and limited data processing capabilities, especially when multiple distributed RISs are deployed to further enhance the coverage and data rate performance. The aim of this paper is to design a scalable and distributed framework for the joint optimization of user scheduling and precoding in distributed RIS-aided multi-user MISO (MU-MISO) systems.
	However, with the increasing number of RISs and users, optimization-based methods lead to significant computational complexity, and the centralized processing incurs significant signaling overhead. 
	While the introduction of DRL algorithms is able to reduce the computational complexity, the performance of single-agent DRL methods is limited due to the curse of dimensionality. Therefore, we propose a scalable and generalizable MADRL framework to overcome the curse of dimensionality and ensure the distributed deployment.
	To reduce the channel training overhead and improve the robustness against channel uncertainty, we leverage statistical CSI for the joint optimization.
	
	Despite the above advantages of the MADRL framework, there still exist some technical challenges for its application in distributed RIS-aided systems. 
	Firstly, a critical challenge lies in ensuring that the network's action outputs meet the constraints of the joint optimization without affecting network training. 
	Secondly, the scalability of the MADRL framework is still impeded by the observation dimension. As the number of RISs and users increase, directly incorporating statistical CSI into the observation leads to much larger network dimensionality, ultimately degrading network performance.
	Moreover, the design of initial observations for the MADRL framework is also challenging, where the random initialization impedes the convergence speed of the algorithm.
	To tackle the above challenges, we employ a centralized optimization-based method for initial observations so as to design the observation and action of the MADRL framework accordingly.

	Specifically, the contributions of this paper are summarized as follows:
	\begin{itemize}
		\item{
		For the joint optimization problem, we first propose a joint brute-force search (BFS) and AO method, where an approximation of the ergodic sum rate is derived to obtain the suboptimal solution.
		Specifically, the served users are obtained through BFS, and the joint precoding design is based on fractional programming (FP) and manifold optimization (MO). Moreover, the solution of the first iteration can provide initial observations for the MADRL framework to accelerate network convergence, and the suboptimal solution upon convergence serves as a centralized benchmark to validate the effectiveness of the MADRL framework.
		}
		\item{To reduce the computational complexity, we propose a MADRL framework to jointly optimize user scheduling and precoding for distributed RIS-aided systems. The scheduling and BS precoding agents are responsible for user scheduling and active precoding, while passive precoding is determined by the corresponding RIS agent. 
		For the local action, we propose a user codebook and reshape the network's action outputs to address the constraints on the number of scheduled users and joint precoding, respectively. 
		For the local observation, to reduce the dimension, the product of multiple statistical CSI and the local action is employed in the observation design, which also ensures that the observation dimension remains fixed as the number of RIS elements increases.
		}
		\item{For the extension of the MADRL framework, a novel dynamic working process of the cross-layer optimization is proposed, which consists of three parts: offline centralized training, online channel estimation and online distributed execution. After centralized training at the central processing unit (CPU), RISs are capable of configuring their passive precoding in a distributed manner. In addition, the user scalability and fairness of the MADRL framework are enhanced to better adapt real-world scenarios.
		} 
		\item{Numerical results initially validate the accuracy of the ergodic sum rate approximation and the convergence of the proposed algorithms. Further,	
		the impact of environmental factors on algorithm performance is investigated, including the BS transmit power, the Rician factor, etc. Our simulation findings demonstrate that the proposed MADRL algorithm shows superior performance with low computational complexity and strong scalability to the system parameters.
		}
	\end{itemize}

	\textit{Remark 1:} The preliminary work for this paper has been presented in \cite{wang2023multiagent}. To be specific, a MADDPG algorithm is proposed to optimize the joint precoding for distributed RIS-aided systems. 
	In comparison, user scheduling is further considered in this paper, with
	a BFS-AO method and a MADRL framework for the cross-layer optimization. Moreover, simulation results show that the proposed MADRL framework outperforms the previous MADDPG algorithm.
	
	\subsection{Organization and Notations}
	The rest of the paper is organized as follows. 
	In Section II, we introduce a distributed RIS-aided downlink MU-MISO system, and formulate a cross-layer optimization problem with the aim of maximizing the ergodic sum rate approximation. Afterwards, a BFS-AO method is proposed to provide initial observations for the MADRL algorithm in Section III, and then a MADRL framework is researched in Section IV. Utilizing the centralized training with decentralized execution (CTDE) paradigm and the trained MADRL algorithm, we propose a dynamic working process in Section V. Extensive experiments are conducted in Section VI, and finally, we draw a conclusion in Section VII. 
	
	The notations in this paper are listed as follows. 
	$(\cdot)^*$, $(\cdot)^T$, $(\cdot)^H$ and $(\cdot)^{-1}$ represent conjugate, transpose, conjugate transpose and inverse operators, respectively; $\mathbb{E}[\cdot]$ denotes the expectation operation, and $\operatorname{Re}\{\cdot\}$ means the real part of the argument; $\operatorname{diag}(\cdot)$ and $\operatorname{blkdiag}(\cdot)$ are diagonal and block diagonal operations; $|\cdot|$ and $\|\cdot\|$ represent the absolute value of a scalar and the Euclidean norm of a vector; $\operatorname{Tr}(\cdot)$ means the trace of a matrix; $\binom{a}{b}$ denotes the number of combinations of $b$ elements chosen from $a$ elements; $\mathcal{C N}\left(\mu, \sigma^{2}\right)$ is symmetric complex Gaussian distribution with mean $\mu$ and variance $\sigma^{2}$; $\mathbf{I}_{M}$ denotes a $M \times M$ identity matrix, and $\mathbf{0}_{M}$ is a $M \times M$ zero matrix.
	
	\section{System Model And Problem Formulation}
	\subsection{System Model}
\begin{figure}[htbp]
		\centering
		\includegraphics[width=0.45\textwidth]{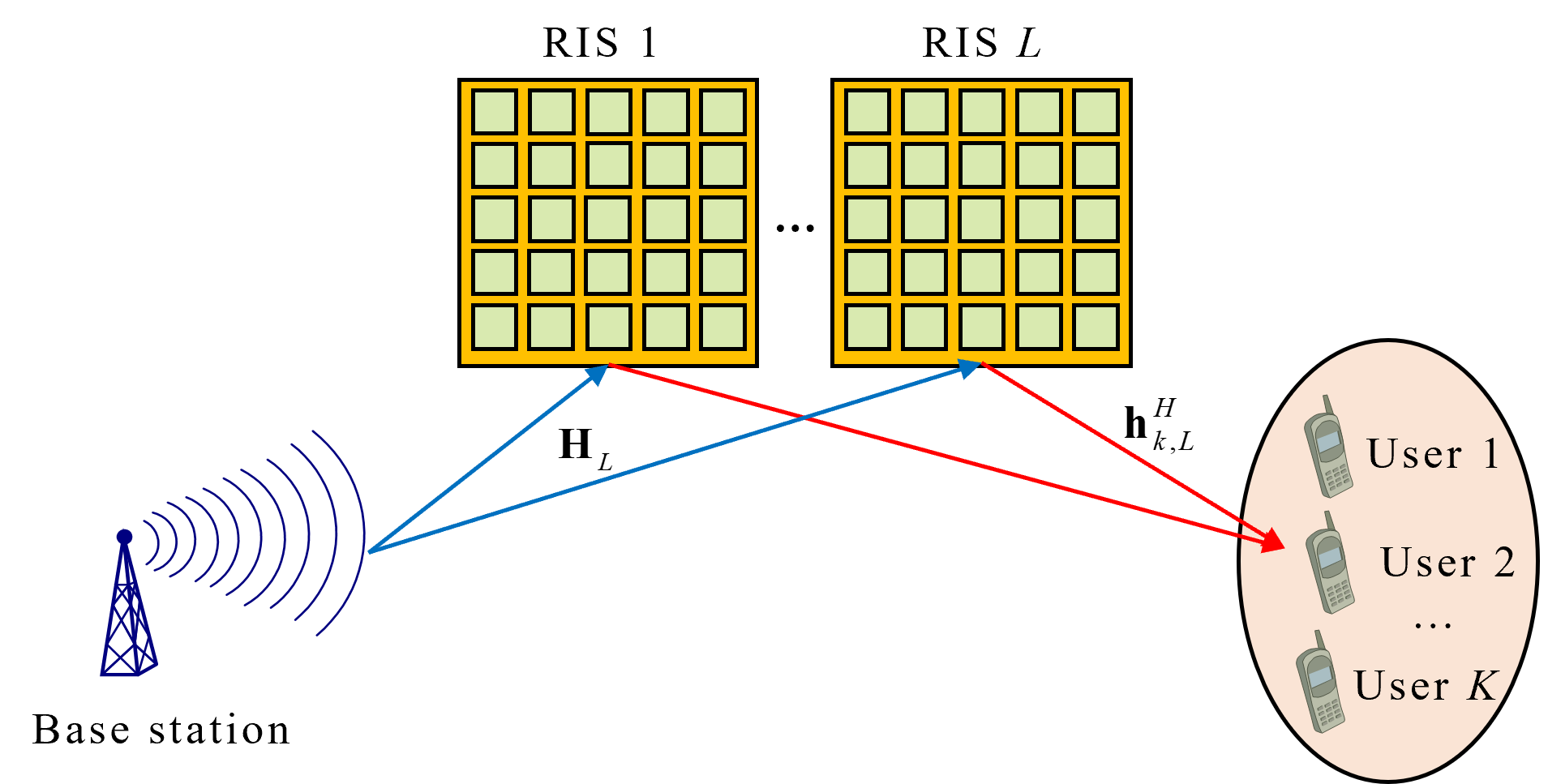}
		\caption{Distributed RIS-aided downlink MU-MISO system.}
		\label{fig1}
	\end{figure}

	We consider a distributed RIS-aided downlink MU-MISO system, as shown in Fig.~\ref{fig1}. It consists of a single BS with $M$ antennas, $K$ single-antenna users and $L$ RISs. The system can serve up to $U$ users ($U < K$) simultaneously, and each of RIS is deployed with $N=N_x \times N_y$ ($N_x$ rows, $N_y$ columns) passive elements.
	We assume that the direct links from the BS to the users are very weak or blocked by obstacles. For tractability, we do not consider the signals reflected multiple times by the RISs. In addition, we also assume perfect statistical CSI available at BS. Let $\mathbf{H}_{l} \in \mathbb{C}^{N \times M}$ denote the channel matrix from the BS to RIS $l$ and define $\mathbf{h}_{k, l}^{H} \in \mathbb{C}^{1 \times N}$ as the channel vector from RIS $l$ to user $k$. The phase shift matrix at RIS $l$ is denoted as 
	$\boldsymbol{\Phi}_{l}=\operatorname{diag}\left(\phi_{l,1}, \phi_{l,2}, \ldots, \phi_{l,N}\right) \in \mathbb{C}^{N \times N}$, where $\phi_{l,n}=e^{j \theta_{l,n}}$ and $\theta_{l,n}$ is the phase shift of the $n$-th element of the $l$-th RIS. The precoding matrix at BS is represented as $\mathbf{G} \in \mathbb{C}^{M \times U}$, and the $k$-th column of this matrix $\mathbf{g}_{k} \in \mathbb{C}^{M \times 1}$ is the precoding vector for scheduled user $k$. Let $\boldsymbol{\alpha} \in \mathbb{C}^{1 \times K}$ denote the scheduling vector, where its $k$-th element $\alpha_k=1$ represents that user $k$ is scheduled, and $\alpha_k=0$ otherwise. The index sets of the RISs, RIS elements, users and scheduled users are written as $\mathcal{L}=\{1, \ldots, L\}$, $\mathcal{N}=\{1, \ldots, N\}$, $\mathcal{K}=\{1, \ldots, K\}$ and $\mathcal{U}=\{k_1, \ldots, k_U\}$, respectively.
	
	For simplicity, the large-scale pass loss of each channel is modeled as \cite{wu2019intelligent}
	\begin{equation} \label{eq1}
		\beta=\beta_0\Bigl(\frac{d}{d_0}\Bigr)^{-\xi},
	\end{equation}
	where $\beta_0$ denotes the path loss at the reference distance $d_0=1$ m; $d$ represents the link distance, and $\xi$ is the path loss exponent. We consider the small-scale fading, and the Rician fading channel model is adopted. Thus, the channels are modeled as
	\begin{align} \label{eq2}
		\mathbf{H}_{l} & =\sqrt{\beta_{l}}\biggl(\sqrt{\frac{K_{l}}{K_{l}+1}} \overline{\mathbf{H}}_{l}+\sqrt{\frac{1}{K_{l}+1}} \tilde{\mathbf{H}}_{l}\biggr),
		\\[0.3mm] \label{eq3}
		\mathbf{h}_{k,l} & =\sqrt{\beta_{k, l}}\biggl(\sqrt{\frac{K_{k, l}}{K_{k, l}+1}} \overline{\mathbf{h}}_{k, l}+\sqrt{\frac{1}{K_{k, l}+1}} \tilde{\mathbf{h}}_{k, l}\biggr),
	\end{align}
	where $K_{l}$ and $K_{k, l}$ denote the Rician factors of $\mathbf{H}_{l}$ and $\mathbf{h}_{k,l}$ respectively; $\beta_{l}$ and $\beta_{k, l}$ are path loss coefficients of the corresponding channels; $\tilde{\mathbf{H}}_{l}$ and $\tilde{\mathbf{h}}_{k, l}$ represent the non-line-of-sight (NLoS) components, each element of which is independent and follows the standard complex Gaussian distribution. $\overline{\mathbf{H}}_{l}$ and $\overline{\mathbf{h}}_{k, l}$ denote the line-of-sight (LoS) components, which can be further expressed as
	\begin{align} \label{eq4}
		\overline{\mathbf{H}}_{l}&=\boldsymbol{a}_{RIS}\left(\varphi_{l}^{A}, \theta_{l}^{A}\right) \boldsymbol{a}_{BS}^{H}\left(\theta_{l}^{D}\right), \\[1mm] \label{eq5}
		\overline{\mathbf{h}}_{k, l}&=\boldsymbol{a}_{RIS}\left(\varphi_{k,l}^{D}, \theta_{k,l}^{D}\right),
	\end{align}
	where
	\begin{gather} \label{eq6}
		\mathbf{a}_{RIS}(\varphi, \theta)=\Bigl[1, \ldots, e^{j \frac{2 \pi}{\lambda} d_{R}(p \sin \varphi \sin \theta+q \cos \theta)}, \ldots, \\
		e^{j \frac{2 \pi}{\lambda} d_{R}((N_x-1) \sin \varphi \sin \theta+(N_y-1) \cos \theta)}\Bigr]^{T},  \nonumber\\ \label{eq7}
		\textbf{a}_{BS}\left(\theta\right)=\Bigl[1,e^{j\frac{2\pi}{\lambda}d_{B}\sin\left(\theta\right)}, \dots, e^{j\frac{2\pi}{\lambda}d_{B} \left(M-1\right) \sin\left(\theta\right)}\Bigr]^T,
	\end{gather}
	where $p=0, \ldots, N_x-1$, $q=0, \ldots, N_y-1$, $\varphi_{l}^{A}$ and $\theta_{l}^{A}$ denote the elevation and azimuth angle of arrival (AoA) at RIS $l$ from BS, respectively; $\varphi_{k,l}^{D}$ and $\theta_{k,l}^{D}$ are the elevation and azimuth angle of departure (AoD) from RIS $l$ to user $k$; $\theta_{l}^{D}$ represents the AoD from BS to RIS $l$; $\lambda$ is the wavelength; $d_{R}$ and $d_{B}$ denote the reflecting element spacing at RIS and the antenna spacing at BS, respectively.
	
	Based on the system model, the received signal of user $k$ can be expressed as
	\begin{equation} \label{eq8}
		y_k=\sum_{l=1}^L \mathbf{h}_{k, l}^H \boldsymbol{\Phi}_l \mathbf{H}_l \alpha_k \mathbf{g}_k x_k + \sum_{l=1}^L \mathbf{h}_{k, l}^H \boldsymbol{\Phi}_l \mathbf{H}_l \!\!\!\sum_{n=1, n \neq k}^K \!\!\!\alpha_n \mathbf{g}_n x_n+w_k,
	\end{equation}
	where $x_{k}$ denotes the transmitted symbol for user $k$ satisfying $\mathbb{E}[|x_{k}|^{2}]=1$, and $w_{k} \sim \mathcal{C N}\left(0, \sigma_{k}^{2}\right)$ is the additive white Gaussian noise at user $k$. For simplicity, we assume that $\sigma_{k}^{2}=\sigma_{0}^{2}$, $\forall k \in \mathcal{K}$. Therefore, the signal-to-interference-plus-noise ratio (SINR) for the $k$-th user is obtained as
	\begin{equation} \label{eq9}
		\rho_k=\frac{\alpha_k\left|\sum_{l=1}^L \mathbf{h}_{k, l}^H \mathbf{\Phi}_l \mathbf{H}_l \mathbf{g}_k\right|^2}{\sigma_0^2+\sum_{n=1, n \neq k}^K \alpha_n\left|\sum_{l=1}^L \mathbf{h}_{k, l}^H \mathbf{\Phi}_l \mathbf{H}_l \mathbf{g}_n\right|^2}.
	\end{equation}
	The ergodic rate of user $k$ can be written as
	\begin{equation} \label{eq10}
		R_{k}=\mathbb{E}\bigl[\log_{2}\left(1+\rho_{k}\right)\bigr].
	\end{equation}
	According to \cite{li2016statistical}, we can obtain the following ergodic rate approximation\footnote{The accuracy of the approximation will be validated in Section \ref{subsec3}.}
	\begin{equation} \label{eq11}
		\hat{R}_{k}=\log_{2}\biggl(1+\frac{\alpha_k \mathbf{g}_{k}^{H} \mathbf{Q}_{k} \mathbf{g}_{k}}{\sigma_{0}^{2}+\sum_{n=1, n \neq k}^{K} \alpha_n \mathbf{g}_{n}^{H} \mathbf{Q}_{k} \mathbf{g}_{n}}\biggr),
	\end{equation}
	where $\mathbf{Q}_k = \mathbb{E}\bigl[\sum_{i=1, j=1}^L \mathbf{H}_i^H \boldsymbol{\Phi}_i^* \mathbf{h}_{k, i} \mathbf{h}_{k, j}^H \boldsymbol{\Phi}_j \mathbf{H}_j\bigr]$.
	Since each element of NLoS components is independent and follows the standard complex Gaussian distribution, the expectation of the product of odd NLOS components, such as $\mathbb{E}\bigl[\tilde{\mathbf{H}}_i^H\bigr]$ and $\mathbb{E}\bigl[\tilde{\mathbf{H}}_i^H \tilde{\mathbf{h}}_{k, i} \tilde{\mathbf{h}}_{k, j}^H \bigr]$, must be zero. Meanwhile, the expectation of the product of two NLOS components has the following property
	\begin{equation} \label{eq12} 
		\mathbb{E}\bigl[\tilde{\mathbf{H}}_i^H \tilde{\mathbf{H}}_j\bigr] = 	
		\begin{cases}
			\mathbf{I}_{M},& i=j\\
			\mathbf{0}_{M},& i\neq j
		\end{cases}.
	\end{equation}
	Thus, after substituting \eqref{eq2} and \eqref{eq3} into $\mathbf{Q}_{k}$, a new expression only about statistical CSI can be obtained in \eqref{eq13},
	\begin{figure*}[htbp]
		\begin{equation} \label{eq13}
			\mathbf{Q}_{k}=\bigl(\overline{\mathbf{h}}_{k}^{equ}\bigr)^{H} 	\overline{\mathbf{h}}_k^{equ} + \sum_{l=1}^L \frac{K_l \beta_l \beta_{k, l} N}{\left(K_l+1\right)\left(K_{k,l}+1\right)} \mathbf{a}_{B S}(\theta_{l}^{D}) \mathbf{a}_{B S}^H(\theta_{l}^{D}) + \sum_{l=1}^L \frac{\beta_l \beta_{k, l} N}{K_l+1} \mathbf{I}_M.
		\end{equation}
		\hrulefill
	\end{figure*}  
	where $\overline{\mathbf{h}}_{k}^{equ}=\overline{\mathbf{h}}_{k}^{H} \boldsymbol{\Phi} \overline{\mathbf{H}}$, $\boldsymbol{\Phi}=\operatorname{blkdiag}\left(\boldsymbol{\Phi}_{1}, \ldots, \boldsymbol{\Phi}_{L}\right)$,
	\begin{align} \label{eq14}
		\overline{\mathbf{h}}_{k}&=\biggl[\sqrt{\frac{\beta_{k,1} K_{k,1}}{K_{k,1}+1}} \overline{\mathbf{h}}_{k, 1}^T, \ldots, \sqrt{\frac{\beta_{k,L} K_{k,L}}{K_{k,L}+1}} \overline{\mathbf{h}}_{k, L}^T\biggr]^T, \\[1mm] \label{eq15}
		\overline{\mathbf{H}}&=\biggl[\sqrt{\frac{\beta_{1} K_{1}}{K_{1}+1}} \overline{\mathbf{H}}_{1}^T, \ldots, \sqrt{\frac{\beta_{L} K_{L}}{K_{L}+1}} \overline{\mathbf{H}}_{L}^T\biggr]^{T}.
	\end{align}

	\subsection{Problem Formulation}
	In this paper, we design the joint optimization of user scheduling and precoding for distributed RIS-aided systems to maximize the ergodic sum rate approximation. 
	Thus, the cross-layer optimization problem is formulated as
	\begin{subequations}
		\begin{align} \label{eq16a}
			\text{(P1)} \quad & \max_{\boldsymbol{\alpha}, \mathbf{G}, \boldsymbol{\Phi}} \sum_{k=1}^K \hat{R}_{k}, \\ \label{eq16b}
			& \hspace{0.05cm} \text { s.t.} \hspace{0.3cm} \sum_{n=1}^K \alpha_n\left\|\mathbf{g}_n\right\|^2 \leq P_{\max}, \\ \label{eq16c}
			& \hspace{0.95cm} \left|\phi_{l,n}\right|=1, \hspace{0.5cm} \forall l \in \mathcal{L}, \forall n \in \mathcal{N},\\ \label{eq16d}
			& \hspace{0.95cm} \alpha_k \in\{0,1\}, \hspace{0.3cm} \forall k\in \mathcal{K},\\ \label{eq16e}
			& \hspace{0.95cm} \sum_{k=1}^K \alpha_k \leq U,
		\end{align}
	\end{subequations}
	where $P_{\max}$ represents the maximum transmission power at BS; \eqref{eq16b} denotes the transmission power constraint at BS; \eqref{eq16c} is the unit power constraint of each RIS element; \eqref{eq16e} indicates that up to $U$ users are scheduled from $K$ users for service.
	Due to the discreteness of the scheduling vector and unit modulus constraints, the cross-layer optimization problem is a mixed-integer nonlinear programming (MINLP) problem. In order to obtain efficient suboptimal solutions of (P1), we first propose a BFS-AO method in Section III, and then, a MADRL based framework is introduced to further reduce the computational complexity in Section IV.
	
	\section{BFS-AO Method}	
	Due to the difficulty of updating the scheduling vector with iterations, traditional AO methods cannot be directly applied to the cross-layer optimization. Therefore, we decompose the optimization problem into two subproblems: user scheduling and joint precoding. In this section, we obtain the optimal scheduling vector via BFS, and design the joint precoding under the fixed scheduling scheme via AO to solve (P1).
	
	\subsection{User Scheduling}
	For the given $\mathbf{G}$ and $\boldsymbol{\Phi}$, we simplify (P1) to the following form 
	\begin{align} \label{eq17}
		\text{(P2)} \quad & \max_{\boldsymbol{\alpha}} \sum_{k=1}^K 	\log_{2}\biggl(1+\frac{\alpha_k \mathbf{g}_{k}^{H} \mathbf{Q}_{k} \mathbf{g}_{k}}{\sigma_{0}^{2}+\sum_{n=1, n \neq k}^{K} \alpha_n \mathbf{g}_{n}^{H} \mathbf{Q}_{k} \mathbf{g}_{n}}\biggr), \nonumber \\
		& \hspace{0.05cm} \text { s.t. } \eqref{eq16d}, \eqref{eq16e}.
	\end{align}
	It can be observed that (P2) is a zero-one integer nonlinear programming problem. 
	Although the computational complexity rapidly increases with the growth of $U$ and $K$, the global optimal solution for $\boldsymbol{\alpha}$ can be obtained by BFS.
	
	\subsection{Joint Precoding}
	For the given $\boldsymbol{\alpha}$, we formulate (P1) to the following form
	\begin{align} \label{eq18}
		\text{(P3)} \quad & \max_{\mathbf{G}, \boldsymbol{\Phi}} \sum_{k=1}^K 	\log_{2}\biggl(1+\frac{\alpha_k \mathbf{g}_{k}^{H} \mathbf{Q}_{k} \mathbf{g}_{k}}{\sigma_{0}^{2}+\sum_{n=1, n \neq k}^{K} \alpha_n \mathbf{g}_{n}^{H} \mathbf{Q}_{k} \mathbf{g}_{n}}\biggr), \nonumber \\ 
		& \hspace{0.05cm} \text { s.t. } \eqref{eq16b}, \eqref{eq16c}.
	\end{align}
	It can be seen that (P3) is a non-convex nonlinear programming problem, which can be solved based on
	FP and MO methods. After applying the Lagrangian dual transform, the new objective function is defined as
	\begin{equation} \label{eq19}
		\max_{\mathbf{G}, \boldsymbol{\Phi}} \sum_{k=1}^K\biggl[\ln \left(1+\varepsilon_k\right)-\varepsilon_k+\frac{\left(1+\varepsilon_k\right) \alpha_k \mathbf{g}_k^H \mathbf{Q}_k \mathbf{g}_k}{\sigma_0^2+\sum_{n=1}^K \alpha_n \mathbf{g}_n^H \mathbf{Q}_k \mathbf{g}_n}\biggr],
	\end{equation}
	where $\boldsymbol{\epsilon}=\left[\epsilon_1, \ldots, \epsilon_K\right] \in \mathbb{C}^{1 \times K}$ is the vector of the auxiliary variables. By taking the first-order derivative, the optimal $\boldsymbol{\epsilon}$ is expressed as
	\begin{equation} \label{eq20}
		\varepsilon_k=\frac{\alpha_k \mathbf{g}_{k}^{H} \mathbf{Q}_{k} \mathbf{g}_{k}}{\sigma_{0}^{2}+\sum_{n=1, n \neq k}^{K} \alpha_n \mathbf{g}_{n}^{H} \mathbf{Q}_{k} \mathbf{g}_{n}}.
	\end{equation}
	Then, for the fixed $\boldsymbol{\epsilon}$, \eqref{eq19} can be further reduced to 
	\begin{equation} \label{eq21}
		\max_{\mathbf{G}, \boldsymbol{\Phi}}
		\sum_{k=1}^K \frac{\left(1+\varepsilon_k\right) \alpha_k \mathbf{g}_k^H \mathbf{Q}_k \mathbf{g}_k}{\sigma_0^2+\sum_{n=1}^K \alpha_n \mathbf{g}_n^H \mathbf{Q}_k \mathbf{g}_n}.
	\end{equation}

	\subsubsection{Active Precoding}
	For the given $\boldsymbol{\epsilon}$ and $\boldsymbol{\Phi}$, we utilize the quadratic transform by introducing $\boldsymbol{\eta} \in \mathbb{C}^{1 \times K}$, $\boldsymbol{\gamma} \in \mathbb{C}^{K \times L}$ and $\boldsymbol{\mu} \in \mathbb{C}^{M \times K}$. The objective function \eqref{eq21} is converted as
	\begin{align} \label{eq22}
		&\max_{\mathbf{G}}
		\sum_{k=1}^{K}\biggl[2 \alpha_k \sqrt{1+\varepsilon_{k}} \operatorname{Re}\left\{\eta_{k}^{*} \overline{\mathbf{h}}_{k}^{e q u} \mathbf{g}_{k}\right\} - \left|\eta_{k}\right|^{2} C_k\biggr] \nonumber\\
		&+\sum_{k=1}^{K} \sum_{i=1}^{L}\biggl[2 \alpha_k \sqrt{1+\varepsilon_{k}} A_{k,i} \operatorname{Re}\left\{\gamma_{k, i}^{*} \mathbf{a}_{B S}^H(\theta_{i}) \mathbf{g}_{k}\right\} - \left|\gamma_{k,i}\right|^{2} C_k\biggr] \nonumber\\
		&+\sum_{k=1}^{K}\biggl[2 \alpha_k \sqrt{1+\varepsilon_{k}} B_k \operatorname{Re}\left\{\boldsymbol{\mu}_{k}^{H} \mathbf{g}_{k}\right\}
		-\left\|\boldsymbol{\mu}_{k}\right\|^{2} C_k \biggr],
	\end{align}
	where $A_{k,i}=\sqrt{\frac{K_i \beta_i \beta_{k, i} N}{\left(K_i+1\right)\left(K_{k, i}+1\right)}}$, $B_k=\sqrt{\sum_{i=1}^L \frac{\beta_i \beta_{k, i} N}{K_i+1}}$ and 
	$C_k=\sigma_{0}^{2}+\sum_{n=1}^{K} \alpha_n \mathbf{g}_{n}^{H} \mathbf{Q}_{k} \mathbf{g}_{n}$; $\eta_{k}$ denotes the $k$-th element of $\boldsymbol{\eta}$; $\gamma_{k, i}$ is the element of $\boldsymbol{\gamma}$ at row $k$ and column $i$, and $\boldsymbol{\mu}_{k}$ means the $k$-th column of $\boldsymbol{\mu}$.
	Through taking the derivative, the optimal auxiliary variables are denoted as
	\begin{align} \label{eq23}
		\eta_k&=\frac{\alpha_k \sqrt{1+\varepsilon_k} \overline{\mathbf{h}}_k^{e q u} \mathbf{g}_k}{\sigma_0^2+\sum_{n=1}^K \alpha_n \mathbf{g}_n^H \mathbf{Q}_k \mathbf{g}_n}, \\[1.5mm] \label{eq24}
		\gamma_{k, i}&=\frac{\alpha_k \sqrt{1+\varepsilon_k} A_{k,i} \mathbf{a}_{B S}^H\left(\theta_i\right) \mathbf{g}_k}{\sigma_0^2+\sum_{n=1}^K \alpha_n \mathbf{g}_n^H \mathbf{Q}_k \mathbf{g}_n}, \\[1.5mm] \label{eq25}
		\boldsymbol{\mu}_k&=\frac{\alpha_k \sqrt{1+\varepsilon_k} B_k \mathbf{g}_k}{\sigma_0^2+\sum_{n=1}^K \alpha_n \mathbf{g}_n^H \mathbf{Q}_k \mathbf{g}_n}.
	\end{align}
	For the fixed auxiliary variables, the optimal $\mathbf{G}$ is obtained in \eqref{eq26}, where $\lambda$ is the dual variable and defined by 
	\begin{figure*}[t]
		\begin{equation} \label{eq26}
			\mathbf{g}_k=\biggl[\lambda \alpha_k \mathbf{I}_M+\sum_{i=1}^K 	\alpha_i\biggl(\left|\eta_i\right|^2+\sum_{j=1}^L\left|\gamma_{i, j}\right|^2+\left\|\boldsymbol{\mu}_i\right\|^2\biggr) \mathbf{H}_i\biggr]^{-1} 
			{\biggl[\alpha_k\sqrt{1+\varepsilon_k}\biggl(\eta_k^*\bigl(\overline{\mathbf{h}}_k^{equ}\bigr)^H + \sum_{j=1}^K A_{k,j} \gamma_{k, j}^* \mathbf{a}_{BS} \left(\theta_j\right)+B_k \boldsymbol{\mu}_k\biggr)\biggr]},
		\end{equation}
		\hrulefill
	\end{figure*}
	\begin{equation} \label{eq27}
		\lambda=\min \biggl\{\lambda \geq 0: \sum_{n=1}^K 	\alpha_n\left\|\mathbf{g}_n\right\|^2 \leq P_{\max }\biggr\}.
	\end{equation}
	Note that the optimal $\lambda$ can be determined efficiently through the bisection search.

	\subsubsection{Passive Precoding}
	For the given $\boldsymbol{\epsilon}$ and $\mathbf{G}$, we utilize the quadratic transform by introducing $\mathbf{x} \in \mathbb{C}^{1 \times K}$, $\mathbf{y} \in \mathbb{C}^{K \times L}$ and $\mathbf{z} \in \mathbb{C}^{M \times K}$. The objective function \eqref{eq21} is changed as
	\begin{align} \label{eq28}
		&\max_{\boldsymbol{\theta}}
		\sum_{k=1}^{K}\biggl[2 \alpha_k \sqrt{1+\varepsilon_{k}} \operatorname{Re}\left\{x_{k}^{*} \boldsymbol{\theta}^H \mathbf{a}_{k,k}\right\} - \left|x_{k}\right|^{2} C_k\biggr] \nonumber\\[0.1mm]
		&+\sum_{k=1}^{K} \sum_{i=1}^{L}\biggl[2 \alpha_k \sqrt{1+\varepsilon_{k}} A_{k,i} \operatorname{Re}\left\{y_{k, i}^{*} \mathbf{a}_{B S}^H(\theta_{i}) \mathbf{g}_{k}\right\} - \left|y_{k,i}\right|^{2} C_k\biggr] \nonumber\\[0.1mm]
		&+\sum_{k=1}^{K}\biggl[2 \alpha_k \sqrt{1+\varepsilon_{k}} B_k \operatorname{Re}\left\{\mathbf{z}_{k}^{H} \mathbf{g}_{k}\right\}
		-\left\|\mathbf{z}_{k}\right\|^{2} C_k \biggr],
	\end{align}
	where $\boldsymbol{\theta}=\operatorname{diag} \bigl(\boldsymbol{\Phi}^H\bigr)$ and $\mathbf{a}_{i,k} = \operatorname{diag} \left(\overline{\mathbf{h}}_k\right) \overline{\mathbf{H}} \mathbf{g}_i$; $x_{k}$ means the $k$-th element of $\mathbf{x}$; $y_{k, i}$ represents the element of $\mathbf{y}$ at row $k$ and column $i$, and $\mathbf{z}_{k}$ denotes the $k$-th column of $\mathbf{z}$.	
	Through taking the derivative, the optimal auxiliary variables are represented as
	\begin{align} \label{eq29}
		x_k&=\frac{\alpha_k \sqrt{1+\varepsilon_k} \boldsymbol{\theta}^H \mathbf{a}_{k,k}}{\sigma_0^2+\sum_{n=1}^K \alpha_n \mathbf{g}_n^H \mathbf{Q}_k \mathbf{g}_n}, \\[1.5mm] \label{eq30}
		y_{k, i}&=\frac{\alpha_k \sqrt{1+\varepsilon_k} A_{k,i} \mathbf{a}_{B S}^H\left(\theta_i\right) \mathbf{g}_k}{\sigma_0^2+\sum_{n=1}^K \alpha_n \mathbf{g}_n^H \mathbf{Q}_k \mathbf{g}_n}, \\[1.5mm] \label{eq31}
		\mathbf{z}_k&=\frac{\alpha_k \sqrt{1+\varepsilon_k} B_k \mathbf{g}_k}{\sigma_0^2+\sum_{n=1}^K \alpha_n \mathbf{g}_n^H \mathbf{Q}_k \mathbf{g}_n}.
	\end{align}
	After removing constant terms and fixing auxiliary variables, the objective function is formulated as
	\begin{equation} \label{eq32}
		\min_{\boldsymbol{\theta}} \hspace{0.1cm} \boldsymbol{\theta}^H \boldsymbol{U} \boldsymbol{\theta}-2 \operatorname{Re}\bigl\{\boldsymbol{\theta}^H \boldsymbol{v}\bigr\},
	\end{equation}
	where
	\begin{align} \label{eq33}
		\boldsymbol{U}&=\sum_{k=1}^K\biggl(\left|x_k\right|^2+\sum_{i=1}^L\left|y_{k, i}\right|^2+\left\|\mathbf{z}_k\right\|^2\biggr) \sum_{n=1}^K \alpha_k \mathbf{a}_{n, k} \mathbf{a}_{n, k}^H, \\[0.1mm] \label{eq34}
		\boldsymbol{v}&=\sum_{k=1}^K \alpha_k \sqrt{1+\varepsilon_k} x_k^* \mathbf{a}_{k, k}.
	\end{align}
	Due to the manifold constraint of \eqref{eq16c}, we obtain the optimal $\boldsymbol{\theta}$ via the MO method, and its specific steps have been shown in \cite{luo2021reconfigurable}. Finally, the proposed BFS-AO method is summarized in Algorithm \ref{alg1}.
	
	\begin{algorithm}[htbp]
		\caption{The BFS-AO Method}
		\label{alg1}
		\renewcommand{\algorithmicrequire}{\textbf{Input:}}
		\renewcommand{\algorithmicensure}{\textbf{Output:}}
		\begin{algorithmic}[1]
			\REQUIRE $\overline{\mathbf{H}}$ and $\overline{\mathbf{h}}_{k}$, $\forall k\in \mathcal{K}$.
			\ENSURE The optimal $\boldsymbol{\alpha}$, $\mathbf{G}$ and $\boldsymbol{\Phi}$.
			
			\REPEAT 
			\STATE Initialize variables $\mathbf{G}$ and $\boldsymbol{\Phi}$ randomly;
			\REPEAT 
			\STATE Update $\boldsymbol{\varepsilon}$ via \eqref{eq20};
			\STATE Update $\boldsymbol{\eta}$, $\boldsymbol{\gamma}$ and $\boldsymbol{\mu}$ via
			\eqref{eq23}, \eqref{eq24} and \eqref{eq25};
			\STATE Update $\mathbf{G}$ via \eqref{eq26};
			\STATE Update $\mathbf{x}$, $\mathbf{y}$ and $\mathbf{z}$ via
			\eqref{eq29}, \eqref{eq30} and \eqref{eq31};
			\STATE Update $\boldsymbol{\theta}$ via the MO method;
			\STATE $\boldsymbol{\Phi}=\operatorname{diag}\bigl(\boldsymbol{\theta}^H\bigr)$;
			\UNTIL{The objective function of (P3) converges.}
			
			\UNTIL{All possibilities of $\boldsymbol{\alpha}$ are exhausted.}
		\end{algorithmic}
	\end{algorithm}
	\vspace{-0.2em}
	
	\subsection{Convergence and Computational Complexity}
	According to \cite{luo2021reconfigurable}, the proposed AO method is monotonically undiminished over the iterations. Hence, Algorithm \ref{alg1} can be guaranteed to converge at least to a local optimal solution of (P1). In addition, the complexity of optimizing $\boldsymbol{\alpha}$, $\mathbf{G}$ and $\boldsymbol{\Phi}$ are $\mathcal{O}\bigl(\binom{K}{U}\bigr)$, $\mathcal{O}\left(K M^3\right)$ and $\mathcal{O}\left( K^2 L^2 N^2\right)$, respectively. Thus, the computational complexity of the BFS-AO method is obtained as $\mathcal{O}\bigl(I_1 \binom{K}{U}\bigl(I_2 K M^3 + I_3 K^2 L^2 N^2\bigr)\bigr)$, where $I_1$, $I_2$ and $I_3$ denote the iteration times of AO, the bisection search and the MO method, respectively.
	
	\section{MADRL Framework}
	In this section, we first give a general overview of MADRL, followed by a detailed explanation of the multi-agent proximal policy optimization (MAPPO) algorithm. Then, the MAPPO algorithm is applied to the cross-layer optimization. 
	
	\subsection{Overview of MADRL}	
	The MADRL algorithms can be generally summarized into three categories: decentralized training with decentralized execution (DTDE), centralized training with centralized execution (CTCE) and CTDE. Without considering the non-stationarity of the environment, DTDE algorithms train the agents independently by regarding other agents as part of the environment. The representative algorithms of the DTDE paradigm are independent deep Q-networks (IDQN) \cite{tampuu2017multiagent} and independent PPO (IPPO)\cite{witt2020is}.
	Moreover, CTCE algorithms utilize the global observation for agents' training and integrate the inter-agent communication during execution. The popular CTCE algorithms include communication neural net (CommNet) \cite{sukhbaatar2016learning} and ATOC \cite{jiang2018learning}.
	In addition, CTDE is the most common MADRL paradigm, which generally consists of actor-critic algorithms and value-decomposition algorithms.
	For actor-critic algorithms, a centralized critic is used to evaluate the global observation and joint action, while decentralized actors execute the local action based on local observations. The characteristic algorithms with actor-critic structure contain MAPPO\cite{yu2022the} and counterfactual multi-agent (COMA) policy gradients \cite{foerster2018counterfactual}. 
	For value decomposition algorithms, a global Q-function is decomposed into a combination of local Q-functions to guide agents' training. The typical algorithms based on value decomposition include value-decomposition networks (VDN)\cite{sunehag2017valuedecomposition} and QMIX \cite{rashid2020monotonic}, where the global Q-function is the linear sum and monotonic function of the local Q-functions, respectively. 
	Particularly for homogeneous agents, parameter sharing \cite{christianos2021scaling} can improve the cooperation efficiency among agents. 
	
	Based on the above analysis, to overcome the non-stationarity of the environment and enable the distributed execution of the algorithm, the CTDE paradigm is taken into consideration. Further, in terms of algorithm stability and the number of hyper-parameters, MAPPO outperforms other MADRL algorithms, and can deal with both continuous and discrete action space. Therefore, a MAPPO based framework is applied to the cross-layer optimization for distributed RIS-aided systems. 
	
	\subsection{Multi-Agent PPO}
	The MADRL algorithm is based on decentralized partially observable Markov Decision Processes (DEC-POMDP) defined by $\langle \mathcal{S}, \mathcal{A}, O, R, P, n, \gamma\rangle$, where $\mathcal{S}$, $\mathcal{A}$ and $\mathcal{O}$ are the space of state, action and observation; the joint action $\mathbf{a}=\left(a_1, \ldots, a_n\right) \in \mathcal{A}$ and the global observation $\mathbf{o}=\left(o_1, \ldots, o_n\right) \in O $ consist of all the local actions and local observations; $a_i$ and $o_i$ represent the local action and local observation for agent $i$; $R$ denotes the shared reward, and $P$ is the state transition probability given the current state and the joint action; $n$ and $\gamma$ represent the number of agents and the discount factor, respectively.
	
	The MAPPO algorithm is composed of multiple PPO agents \cite{schulman2017proximal} and adopts the CTDE paradigm to make the environment stationary. Due to the fully cooperative relationship among agents, a centralized critic network $V_\phi\left(\mathbf{o}\right)$ with parameters $\phi$ is introduced and the reward $r$ is shared by all agents. 
	For agent $i$, it contains an actor network $\pi_{\theta_i}\left(a_i \mid o_i\right)$ with parameters $\theta_i$, which are shared with homogeneous agents. 
	During time step $t$, each agent obtains the local observation and simultaneously outputs the local action. 	 
	After the joint action is executed, the environment goes to a new state and the shared reward is obtained. Then, a replay buffer $\mathcal{D}$ is applied to store the transitions of all agents, which are defined as
	$\bigl\{\mathbf{o}_t,  \mathbf{a}_t, \mathbf{o}_{t+1}, r_t\bigr\}$, where $\mathbf{o}_t$,  $\mathbf{a}_t$ and $r_t$ denote the global observation, the joint action and the shared reward at time step $t$.
	Afterwards, the optimizer selects a minibatch of transitions from $\mathcal{D}$ to update the networks. The actor network $\pi_{\theta_i}$ is updated through gradient
	ascent to maximize the objective function, which can
	be written as: 
	\begin{equation} \label{eq35}
		J(\theta_i)=\mathbb{E}\bigl[\min \left(r_t(\theta_i) A_t, \operatorname{clip}\left(r_t(\theta_i), 1-\epsilon, 1+\epsilon\right) A_t\right)\bigr],
 	\end{equation}
    where $r_t\left(\theta_i\right)=\frac{\pi_{\theta_i}\left(a_{i,t} \mid o_{i,t}\right)}{\pi_{\theta_{i'}}\left(a_{i,t} \mid o_{i,t}\right)}$ denotes the probability ratio at time step $t$, and $\pi_{\theta_{i'}}$ is the actor network before the update; $a_{i,t}$ and $o_{i,t}$ represent the local action and local observation for agent $i$ at time step $t$, respectively;
	$\operatorname{clip}\left(r_t(\theta), 1-\epsilon, 1+\epsilon\right)$ means that the probability ratio is limited within the range $\left[1-\epsilon, 1+\epsilon\right]$, and $\epsilon$ is the clip parameter.
	To balance the estimated bias and variance, general advantage estimation (GAE) is applied to calculate the advantage function, which can be obtained as
	\begin{gather} \label{eq36}
	A_t=\sum_{l=0}^{\infty} (\gamma \lambda)^l \delta_{t+l}, \\ \label{eq37}
	\delta_t= r_t + \gamma V_\phi\left(\mathbf{o}_{t+1}\right)- V_\phi\left(\mathbf{o}_t \right), 
	\end{gather}
	where $\lambda$ represents the GAE coefficient, and $r_t$ is the reward at time step $t$.
	Through gradient descent, the centralized critic network $V_\phi$ is updated to minimize the loss, which can be expressed as
	\begin{equation} \label{eq38}
	L(\phi)=\mathbb{E}\bigl[\bigl(V_\phi\left(\mathbf{o}_t\right)-A_t-V_{\phi'}\left(\mathbf{o}_t\right)\bigr)^2\bigr],
	\end{equation}
	where $V_{\phi'}$ denotes the critic network before the update. After the centralized training, the actor network of each agent can be deployed in a distributed manner.

	\subsection{Cross-layer Optimization Using MAPPO}  \label{subsec1}
\begin{figure*}[htbp]
		\centering
		\includegraphics[width=0.95\textwidth]{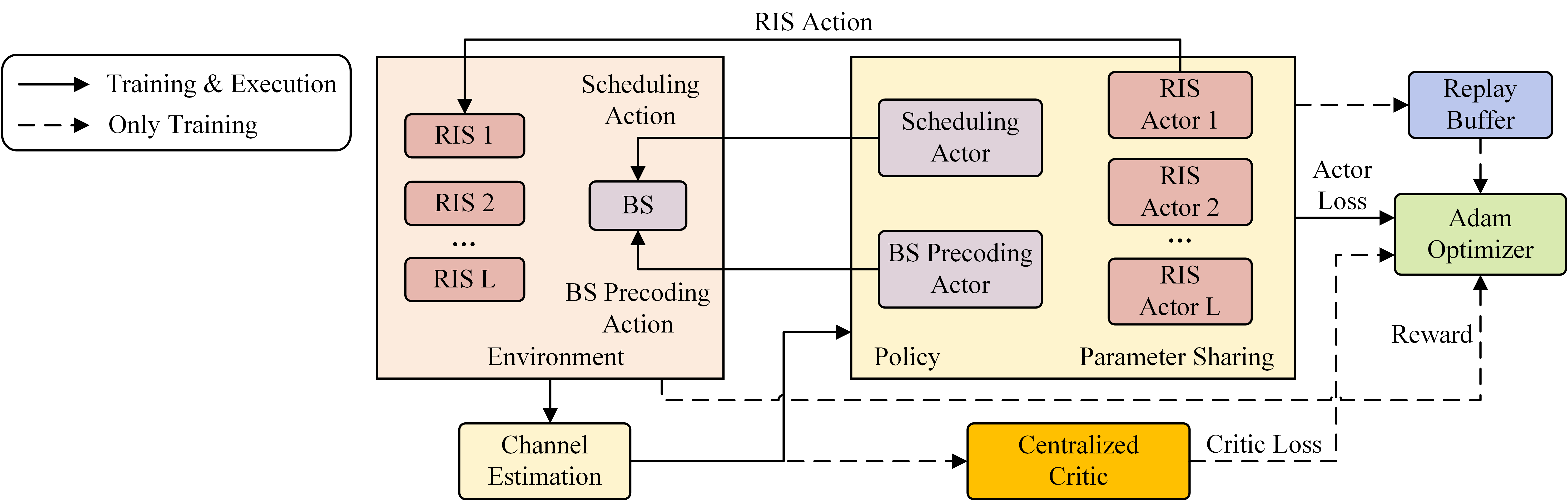}
		\caption{Structure of the proposed MADRL framework.}
		\label{fig2}
	\end{figure*}

	To tackle the problem (P1), we propose a MAPPO algorithm including three types of agents, where the scheduling agent, BS precoding agent and RIS agents are responsible for optimizing user scheduling, active precoding and passive precoding, respectively. 
	For discrete action space (referred to as discrete PPO), the action is selected according to the probability distribution produced by the actor network. In continuous action space (known as continuous PPO), the actor network outputs the mean and variance of a Gaussian distribution, from which the action is sampled.
	The structure of the proposed framework is shown in Fig.~\ref{fig2}, and then important components are defined as follows. Note that the real and imaginary components of each local observation are fed separately into the corresponding actor network.
	
	\subsubsection{Scheduling Agent}
	The scheduling agent employs a discrete PPO to design user scheduling. At time step $t$, the local observation of the scheduling agent is defined as\footnote{The matrices or vectors with the notation $(\cdot)(t)$ represent the corresponding matrices or vectors at time step $t$.}
	\begin{equation} \label{eq39}
		o_{1,t}=\overline{\mathbf{h}}^{H}{(t)} \boldsymbol{\Phi}{(t-1)} \overline{\mathbf{H}}{(t)},
	\end{equation}
	where $\overline{\mathbf{h}}=\bigl[\overline{\mathbf{h}}_{1}, \ldots, \overline{\mathbf{h}}_{K}\bigr] \in \mathbb{C}^{NL \times K}$. 
	
	The action space of the scheduling agent contains all the possible ways to schedule $U$ from $K$ users. 
	Considering the specific case where $K=4$ and $U=2$, the action space can be represented as a codebook with six codewords, which is denoted as $[1,2; 1,3; 1,4; 2,3; 2,4; 3,4]$. The codebook enables a fixed number of scheduled users and eliminates the possibility of duplicated scheduling.
	After the probability distribution of action space is output by the actor network, the index of the codeword with the highest probability is served as the local action. 
	Thus, the dimensions of the local observation and action space are $2KM$ and $\binom{K}{U}$, respectively.

	\subsubsection{BS Precoding Agent}
	The BS precoding agent utilizes a continuous PPO to optimize active precoding. At time step $t$, the local observation of the BS precoding agent is denoted as
	\begin{equation} \label{eq40}
		o_{2,t}=\hat{\mathbf{h}}^{H}{(t)} \boldsymbol{\Phi}{(t-1)} \overline{\mathbf{H}}{(t)}\mathbf{G}{(t-1)},
	\end{equation}
	where $\hat{\mathbf{h}}=\bigl[\overline{\mathbf{h}}_{k_1}, \ldots, \overline{\mathbf{h}}_{k_U}\bigr] \in \mathbb{C}^{NL \times U}$.
	
	The local action of the BS precoding agent at time step $t$ is expressed as
	\begin{equation} \label{eq41}
		a_{2,t}=\mathbf{G}{(t)}.
	\end{equation}
	The real and imaginary components of the local action require to be reformulated into a complex-valued form.
	When generating the local observation at the next time step, $\mathbf{G}{(t)}$ is directly applied. During reward computation, the active precoding $\mathbf{G}(t)=\frac{\sqrt{P_{max}} \mathbf{G}(t)} {\left\|\mathbf{G}(t) \right\|}$ is scaled to satisfy \eqref{eq16b}.
	Therefore, the dimensions of the local observation and local action are $2U^2$ and $2MU$, respectively.

	\subsubsection{RIS Agents}	
	The actor networks of different RIS agents share a set of parameters. For RIS $l$, the RIS agent applies a continuous PPO to optimize passive precoding. At time step $t$, the local observation of RIS agent $l$ is determined by
	\begin{equation} \label{eq42}
		o_{l+2,t}=\overline{\mathbf{h}}_l^{H}{(t)} \boldsymbol{\Phi}_l{(t-1)} \overline{\mathbf{H}}_l{(t)},
	\end{equation}
	where $\overline{\mathbf{h}}_l=\bigl[\sqrt{\frac{\beta_{k_1,l} K_{k_1,l}} {K_{k_1,l}+1}} \overline{\mathbf{h}}_{k_1,l}, \ldots, \sqrt{\frac{\beta_{k_U,l} K_{k_U,l}}{K_{k_U,l}+1}} \overline {\mathbf{h}}_ {k_U,l}\bigr]$. 
	
	The local action of RIS agent $l$ at time step $t$ is denoted as
	\begin{equation} \label{eq43}
		a_{l+2,t}=\left\{\theta_{l,1}(t), \cdots, \theta_{l,N}(t)\right\}.
	\end{equation}
	For purposes of dimension reduction, the local action consists of $N$ real-valued phase shifts instead of a complex-valued phase shift matrix. When obtaining the local observation at the next time step, the local action is directly employed. During reward computation, $\phi_{l,n}=e^{j \theta_{l,n}}$ is obtained and meets \eqref{eq16c}.
	Hence, the dimensions of the local observation and local action are $2MU$ and $N$, respectively.
	
	\subsubsection{Shared Reward}
	Due to the fully cooperative relationship among agents, the reward shared by all agents is the ergodic sum rate approximation. At time step $t$, it is written as
	\begin{equation} \label{eq44}
		r_{t}=\sum_{k=1}^K \hat{R}_{k}(t).
	\end{equation}
	To avoid duplication of information, the global observation only comprises the local observations of the scheduling and BS precoding agent, where all the information observed by RIS agents is strategically included. Thereby, the dimension of the global observation is $2KM+2U^2$.

	\textit{Remark 2:} The global and local observations designed above have the following advantages. Firstly, compared to directly incorporate statistical CSI into the observation, we employ the product of multiple statistical CSI and the local action in the design.
	For the scheduling agent, the previous design incurs an observation dimension of $2NLK+2NL+2NLM$, while the proposed design significantly reduces the dimension to
	$2KM$.
	Secondly, the observation dimension remains fixed as $N$ and $L$ increase, which enhances the scalability of the proposed algorithm.
	Moreover, the elements of each observation matrix have the same order of magnitude, which accelerates the algorithm convergence.
	
	In order to further improve the algorithm performance, we apply some skills in the detailed implementation of MAPPO, including the state normalization, advantage normalization, orthogonal network initialization and so on. 
	To accelerate the training process, the solution from the first iteration of the BFS-AO method provides initial observations for all agents. 
	In the testing phase, observations are initialized by the local action from the last training step.
	Let $\mathcal{I}=\{1,2, \ldots, L+2\}$ denote the index set of all agents, and
	the details of the proposed MAPPO algorithm are presented in Algorithm \ref{alg2}.
	After offline centralized training at BS, distributed RISs can independently configure their own phase shift matrices during online distributed execution.
	
	\begin{algorithm}[htbp]
		\caption{The MAPPO Based Framework}
		\label{alg2}
		\renewcommand{\algorithmicrequire}{\textbf{Input:}}
		\renewcommand{\algorithmicensure}{\textbf{Output:}}
		
		\begin{algorithmic}[1]
			\REQUIRE $\overline{\mathbf{H}}$ and $\overline{\mathbf{h}}$.
			\ENSURE The optimal $\boldsymbol{\alpha}$, $\mathbf{G}$ and $\boldsymbol{\Phi}$.
			
			\STATE \textbf{Initialization:} A centralized critic network $V_{\phi}\left(\mathbf{o}\right)$, actor networks ${\pi}_{{\theta}_{i}} \left(a_i \mid o_i\right)$ for $i\in \mathcal{I}$ and a replay buffer $\mathcal{D}$.
			
			\FOR{each episode}
			\STATE Collect the current $\overline{\mathbf{H}}$, $\overline{\mathbf{h}}$ and obtain initial observations;
			
			\FOR{$t=1, 2, \ldots, T$}
			\STATE Obtain action $a_{i,t}$ from the corresponding networks ${\pi}_{{\theta}_{i}}\left(a_i \mid o_i\right)$ for $i\in \mathcal{I}$;
			\STATE Calculate the reward $r_t$ and collect the current $\overline{\mathbf{H}}$, $\overline{\mathbf{h}}$ to obtain new observations $o_{i,t+1}$ for $i\in \mathcal{I}$;
			\STATE Store a transition $\bigl\{\mathbf{o}_t, \mathbf{a}_t, \mathbf{o}_{t+1}, r_t\bigr\}$ in $\mathcal{D}$;
			\STATE Update observations $o_{i,t}=o_{i,t+1}$ for $i\in \mathcal{I}$;
			\ENDFOR
			\STATE Calculate the advantage function $A_t$;
		
			\FOR{each epoch}
			\REPEAT
			\STATE Sample a minibatch of transitions from $\mathcal{D}$;
			\STATE Update ${\pi}_{{\theta}_{i}}\left(a_i \mid o_i\right)$ for $i\in \mathcal{I}$ through gradient ascent via \eqref{eq35};
			\STATE Update $V_{\phi}\left(\mathbf{o}\right)$ through gradient descent via \eqref{eq38};
			\UNTIL All transitions in $\mathcal{D}$ have been sampled.
			
			\ENDFOR
			\STATE Clear all the transitions in $\mathcal{D}$.
			
			\ENDFOR
		\end{algorithmic}
	\end{algorithm}
	
	\subsection{Computational Complexity}
	Given a fully connected layer with the input dimension $I$ and the output dimension $O$, the complexity is expressed as $\mathcal{O}\left(I O\right)$. Therefore, the computational complexity of the MAPPO algorithm is $\mathcal{O}\bigl(\bigl(KM+MUL+NL+\binom{K}{U}\bigr)I + LI^2\bigr)$, where $I$ denotes the number of neurons in the hidden layer.

	\section{Extension of the MAPPO Algorithm}
	In this section, we develop a dynamic working process by utilizing the CTDE paradigm and the trained algorithm. Then, the user scalability and fairness of the algorithm are enhanced to better adapt real-world scenarios.
	\subsection{Dynamic Working Process}
\begin{figure}[htbp]
		\centering
		\includegraphics[width=0.45\textwidth]{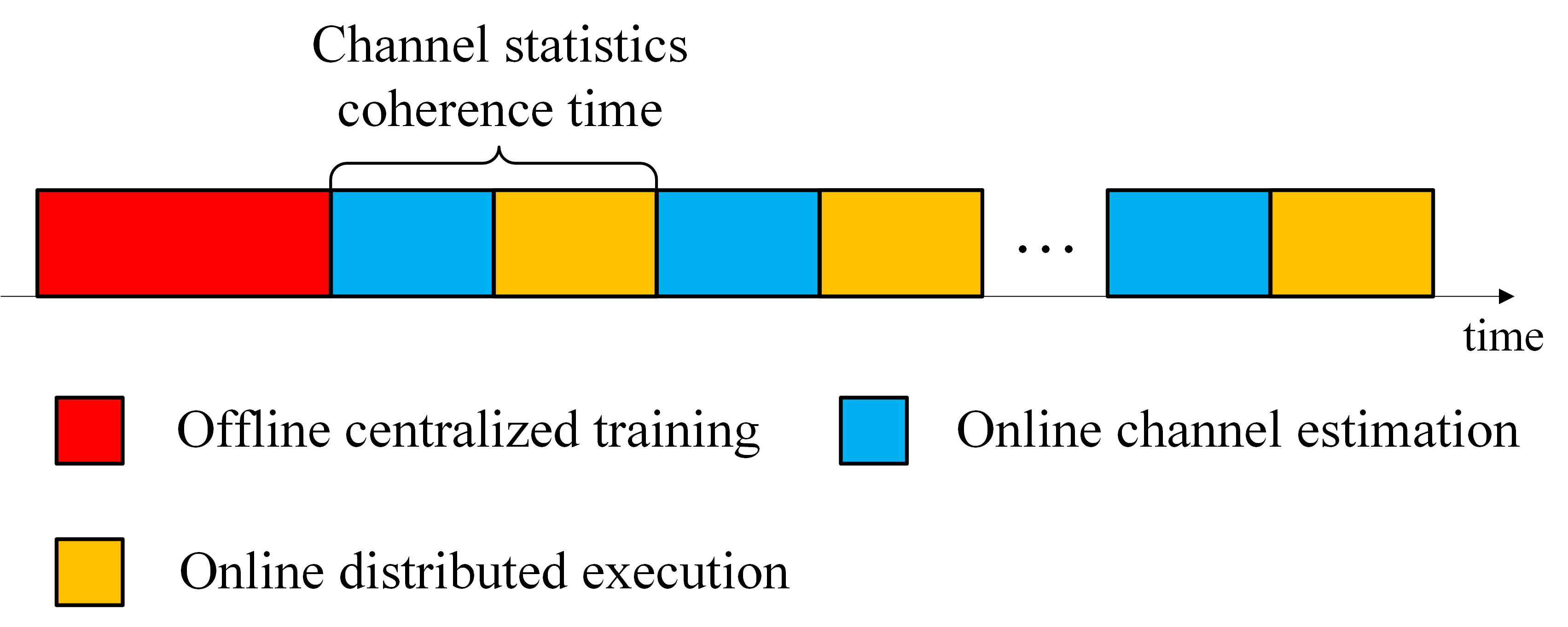}
		\caption{Dynamic working process of the cross-layer optimization.}
		\label{fig3}
	\end{figure}

	In practical scenarios, centralized deployment of algorithms concentrates all computing tasks and data on CPU at BS, which leads to a decline in system performance. To overcome the above limitations, we propose a dynamic working process that leverages distributed RISs to facilitate the distributed deployment of MADRL algorithms\footnote{The sensing RIS architecture \cite{taha2021enabling} can be utilized to achieve the distributed dynamic working process, wherein active sensors are capable of both the channel estimation and signal reflection.}. 
	As illustrated in Fig.~\ref{fig3}, the dynamic working process of the cross-layer optimization consists of three parts, including offline centralized training, online channel estimation and online distributed execution. 

	In the offline centralized training stage, the MAPPO algorithm is trained at the CPU. Then, we deploy the trained actor networks of the BS precoding and scheduling agent at BS, while a shared RIS actor network is arranged at each RIS. At the beginning of channel statistics coherence time, the BS and distributed RISs capture the local observations independently through online channel estimation.	
	Specific channel parameter estimation methods can refer to tensor-based methods \cite{lin2021tensorbased}, \cite{lin2022channel}, which achieve desirable channel estimation accuracy with reduced training overhead. 
	Leveraging the estimated channel parameters, statistical CSI can be obtained from the channel model.
	Next, the local action is obtained according to the corresponding actor network during online distributed execution. As the users' location varies, we continuously perform the channel estimation and distributed execution to complete the cross-layer system configuration. In the next section, it will be demonstrated that the trained networks have strong generalizability to the BS transmit power and transmission environment. Therefore, the networks only require retraining when there are significant changes in the system, and special attention should be given to scenarios where the number of users changes.
	
	\subsection{User Scalability} \label{subsec2}
	The MADRL algorithm, with a fixed total number of users, faces enormous challenges in real-world scenarios where the number of users varies frequently. Due to the unacceptable cost of frequent centralized training, we further improve the scalability of the algorithm. Let $\hat{K}$ denote the actual number of users.
	User scheduling aims to select $U$ out of $K$ users, but we next focus on the situation that $\hat{K}\neq K$. To directly apply the trained algorithm, we either pre-screen the users to $K$ when $\hat{K}>K$, or add virtual users to $K$ before scheduling when $\hat{K}<K$.
	
	For the case that $\hat{K}>K$, we introduce a priority vector into pre-screening to consider user fairness \cite{li2018user}, which is defined as $\mathbf{p}= \left[p_1, \ldots, p_{\hat{K}} \right]$. After each distributed execution,  unscheduled users increase their priorities by one, while the priorities of the scheduled users are set to zero. Before scheduling, we select $K$ users with the highest priority from $\hat{K}$ users, and users with the same priority are selected at random. When $\hat{K}<K$, the total number of users is maintained by adding virtual users with all-zero channel matrices. Then, the optimal action that does not involve the virtual users is chosen by the scheduling agent based on the probabilities of all action. Based on the above extension, the MADRL algorithm can deal with practical scenarios where the number of users varies without retraining.
	
	\subsection{User Fairness}
	When maximizing the ergodic sum rate based on the given users' statistical CSI, some users might not be scheduled. To take user fairness into account, the proposed algorithm can be modified accordingly with Jain’s fairness index (JFI) \cite{sediq2013optimal}, which can be expressed at time step $t$ as
	\begin{equation} \label{eq45}
		JFI_t=\frac{\left(\sum_{k=1}^K \hat{R}_k(t)\right)^2}{K\left(\sum_{k=1}^K \hat{R}_k^2(t)\right)}.
	\end{equation}
	Specifically, a larger JFI value indicates a fairer scheduling algorithm, with $JFI = 1$ representing the ideal scenario where all users achieve the same throughput.
	Then, we can set the shared reward $\hat{r}_t$ at time step $t$ as
	\begin{equation} \label{eq46}
		\hat{r}_t=(1-\upsilon) r_t+ \upsilon J F I_t,
	\end{equation}
	where $\upsilon$ is the fairness weight factor and limited within the range $[0, 1]$. 
	Note that the MAPPO algorithm with $\upsilon=0$ is equivalent to the proposed algorithm in Section \ref{subsec1} and exclusively prioritizes the ergodic sum rate. Conversely, the MAPPO algorithm with $\upsilon=1$ only focuses on user fairness.
	Therefore, by incorporating JFI into the shared reward, the MAPPO algorithm can be trained across varying $\upsilon$ to favor user fairness.
	
	\section{Numerical Results}
	In this section, we evaluate the performance of the proposed algorithm based on numerical results. We begin by presenting the simulation scenario of the system, 	
	and then validate the accuracy of the ergodic sum rate approximation and the convergence of BFS-AO and MAPPO methods. Next, we adjust environmental factors to verify the generalizability of the MADRL algorithm, and then demonstrate its strong scalability to $N$ and $K$.
	
	\subsection{Simulation Setup}
\begin{figure}[htbp]
		\centering
		\includegraphics[width=0.45\textwidth]{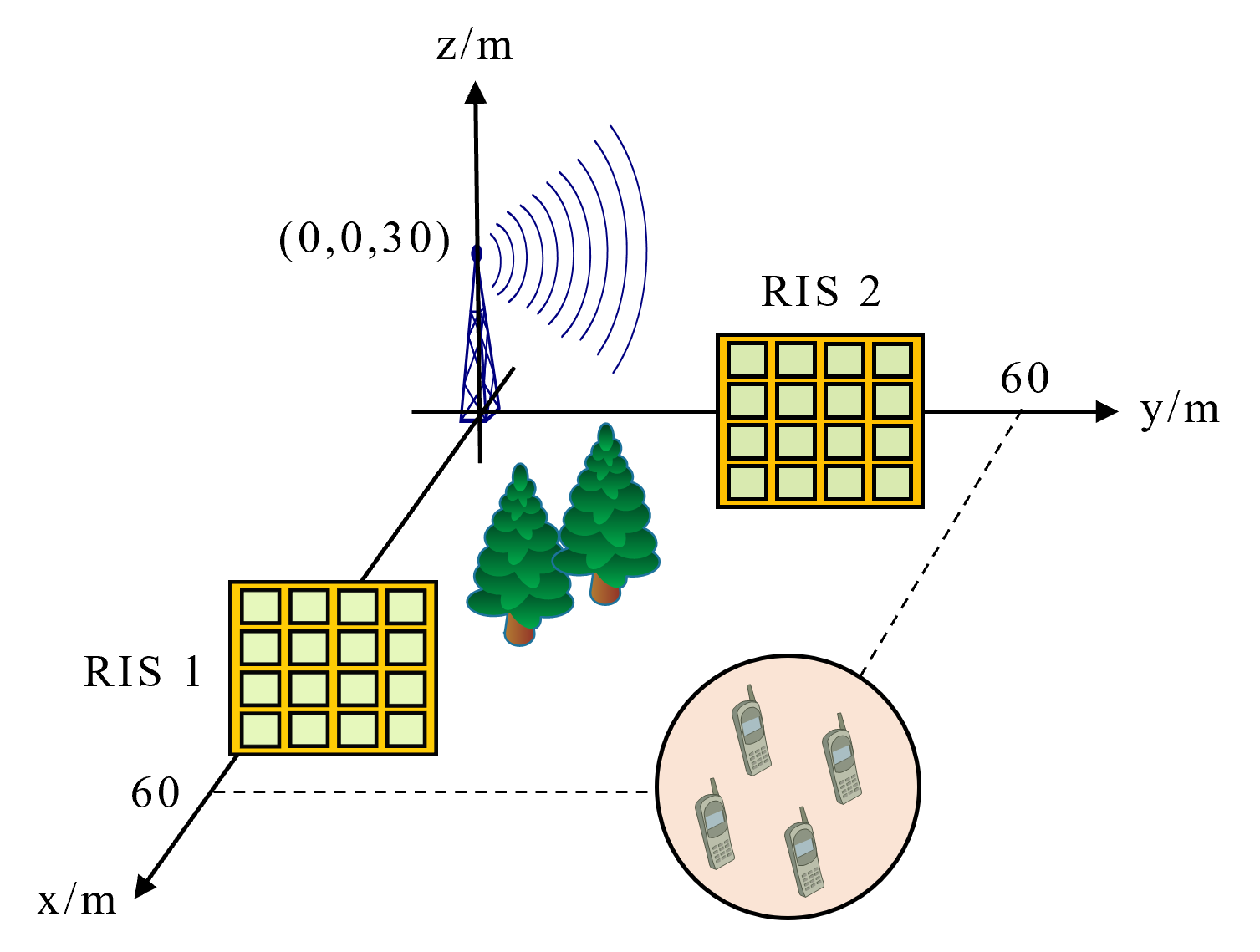}
		\caption{A simulation scenario with two distributed RISs.}
		\label{fig4}
	\end{figure}

	We consider a distributed RIS-aided downlink MU-MISO system as shown in Fig.~\ref{fig4}. In this setup, one BS and two RISs are located at 
	($0$, $0$, $30$ m), ($50$ m, $20$ m, $10$ m) and ($20$ m, $50$ m, $10$ m), respectively. The users randomly lie within a circle centered at ($60$ m, $60$ m, $0$) with radius $r=6$ m. In terms of the large-scale fading, the path loss exponent of the BS-RIS and RIS-User links are expressed as $\xi_0$ and $\xi_1$, and we set $\beta_0=\frac{1}{1000}$, $d_0=1$ m, $\xi_0=2.2$ and $\xi_1=2.8$ \cite{zhang2021joint}. In addition, we set the Rician factor as $K_l=K_{k,l}=6$ dB, $\forall k \in \mathcal{K}, \forall l \in \mathcal{L}$. The maximum transmit power at BS and the noise power are set as $P_{max}=10$ dBm and $\sigma_{0}^{2}=-90$ dBm, respectively. Other system parameters are set as: $M=8$, $N=64$, $K=8$, $U=2$ and $\upsilon=0$. 
	To reduce the impact of randomness on the results, all algorithms are averaged over $300$ independent channel realizations.
	
	For the proposed MAPPO algorithm, all networks are fully connected with one hidden layer, and the detailed hyper-parameters are listed in Table \ref{tab1}. To validate the performance of the MAPPO algorithm, we compare the following algorithms: 
	
	\begin{itemize}
		\item{\textbf{BFS-AO:} Algorithm \ref{alg1} is proposed to serve as a centralized benchmark of the cross-layer optimization.}
		\item{\textbf{BFS-PDS:} The primal-dual subgradient (PDS) method in \cite{zhang2021joint} replaces the passive precoding of the BFS-AO algorithm.}
		\item{\textbf{MAPPO:} Offline training is performed based on Algorithm \ref{alg2}, followed by online execution according to the proposed dynamic working process.}	
		\item{\textbf{BFS-MADDPG:} The optimal scheduling vector is obtained via BFS, while the trained MADDPG method in \cite{wang2023multiagent} is employed to achieve the link-level optimization under each scheduling scheme.}
		\item{\textbf{PPO-MADDPG:} The trained scheduling agent of the MAPPO algorithm is employed, while the MADDPG method is utilized to design the joint precoding.}
		\item{\textbf{PPO-AO:} Considering the learning-optimization framework in \cite{li2024optimal}, the trained scheduling agent and AO method are responsible for user scheduling and joint precoding, respectively.}
		\item{\textbf{Random-MADDPG:} Users are scheduled randomly and the joint  precoding is designed exploiting the MADDPG algorithm.} 
		\item{\textbf{Random precoding:} The random active precoding replaces the active precoding optimization of the BFS-AO algorithm.}
		\item{\textbf{Random RIS:} The random passive precoding replaces the passive precoding optimization of the BFS-AO algorithm.
		}
	\end{itemize}

	\begin{table}[htbp]
		\renewcommand{\arraystretch}{1.25}
		\setlength{\tabcolsep}{6pt}
		\caption{MAPPO Hyper-Parameters}
		\begin{center}
			\scalebox{0.95}{
				\begin{tabular}{l c} 
					\hline\hline
					Parameter&Value\\
					\hline
					Number of training episodes&600\\
					Number of steps in each episode&1024\\
					Replay buffer size&1024\\
					Batch size&128\\
					Sample reuse&10\\
					Discount factor&0.45\\
					GAE coefficient&0.45\\
					Clip parameter&0.3\\
					Entropy coefficient&0.01\\
					Learning rate of all networks&$3e^{-4}$\\
					\hline
					Number of layers for networks&$3$\\
					Hidden layers activation&Tanh\\
					Optimizer&Adam\\
					Final layer activation (scheduling actor)&Softmax\\
					Final layer activation (other networks)&Linear\\
					Number of nodes (scheduling actor)&$(64,28,28)$\\
					Number of nodes (BS precoding actor)&$(16,32,32)$\\
					Number of nodes (RIS actor)&$(32,64,64)$\\
					Number of nodes (centralized critic)&$(64,8,1)$\\
					\hline	
			\end{tabular}}
			\label{tab1}
		\end{center}
	\end{table}
	\vspace{-1.5em}
	
	\subsection{Validation} \label{subsec3}
	\begin{figure}[htbp]
		\centering
		\includegraphics[width=0.45\textwidth]{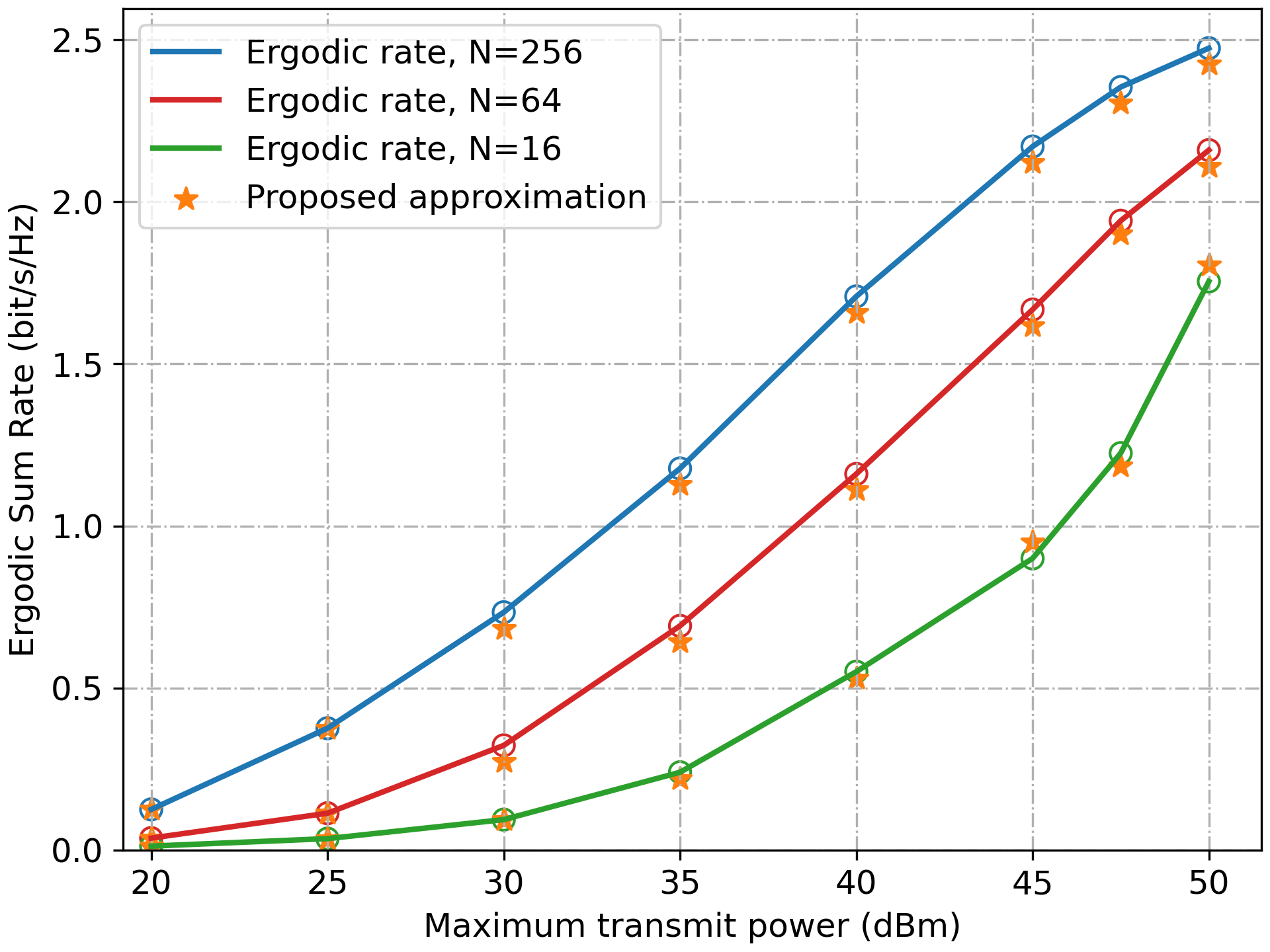}
		\caption{Validation of the ergodic sum rate approximation.}
		\label{fig5}
	\end{figure}

	Fig.~\ref{fig5} validates the accuracy of the ergodic sum rate approximation, where user scheduling and joint precoding are randomly generated. We observe that the closed-form approximation closely tracks the ergodic sum rate as the transmission power increases. In addition, the ergodic rate exhibits a significant improvement with an increasing number of RIS elements. The simulation results show that it is reasonable to design the user scheduling and joint precoding algorithm based on this approximation. Yet, the actual ergodic sum rate is computed via Monte-Carlo methods in the following simulation results for performance comparison.
	
	\subsection{Convergence}
	\begin{figure*}[htbp]
		\centering
		\subfloat[BFS-AO method]{
			\includegraphics[width=0.45\textwidth]{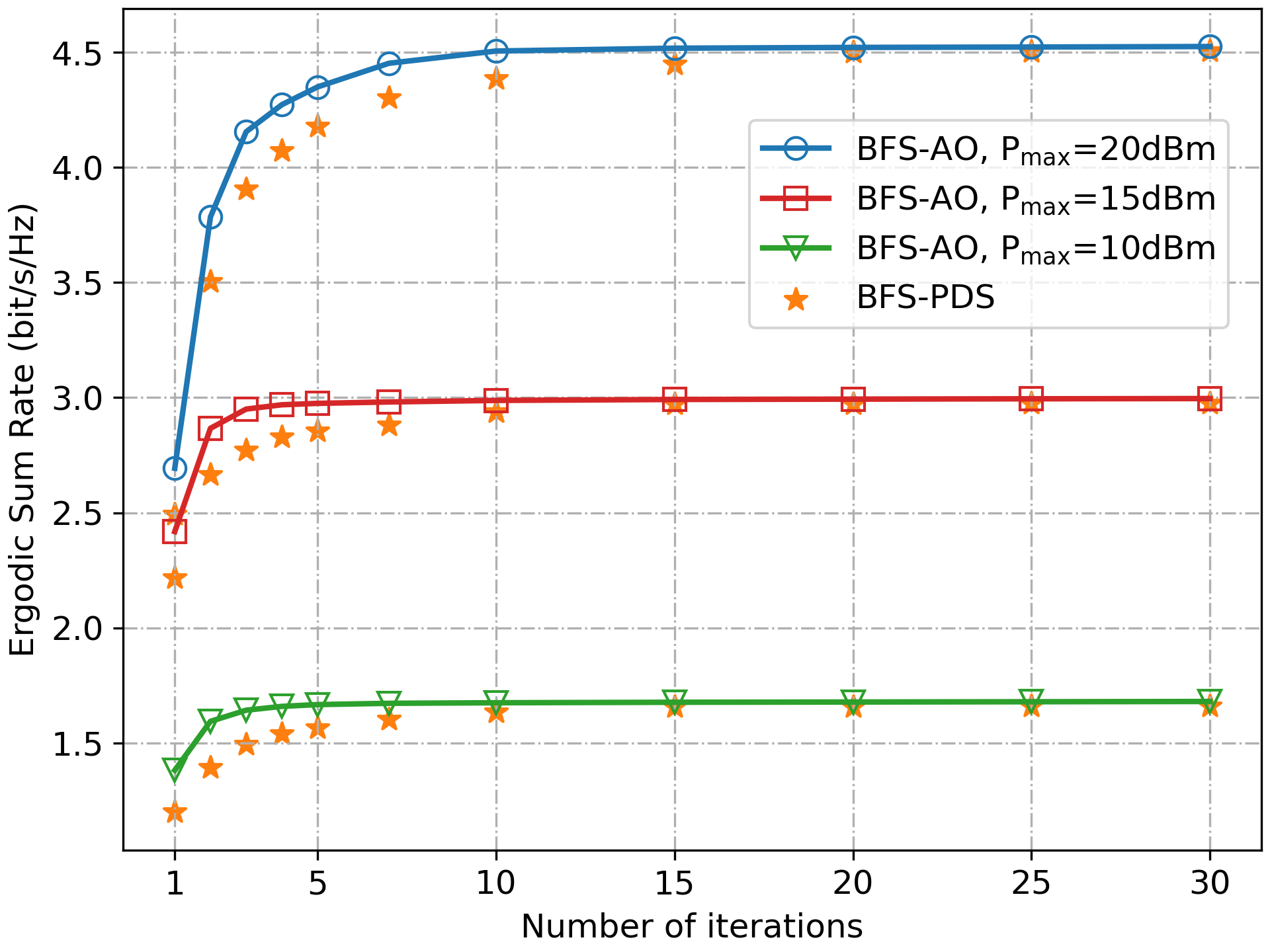}
			\label{fig6(a)}
			\vspace{-0.25cm}}
		\hfill
		\subfloat[MAPPO algorithm]{
			\includegraphics[width=0.45\textwidth]{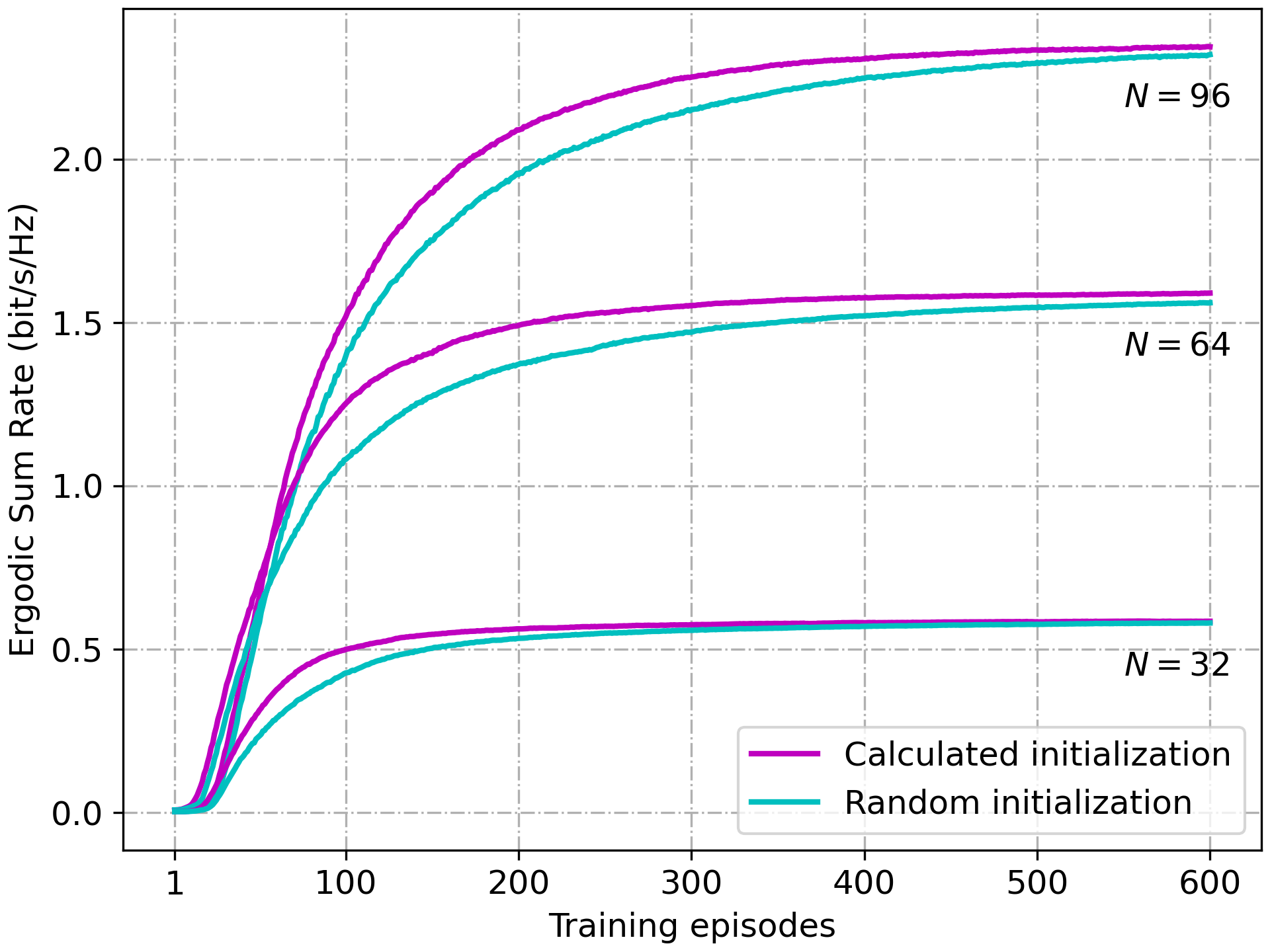}
			\label{fig6(b)}
			\vspace{-0.25cm}}
		\caption{Convergence behavior of the proposed algorithms.}
		\label{fig6}
	\end{figure*}

	Fig.~\ref{fig6} shows the convergence behaviors of two proposed algorithms. 
	In Fig.~\subref*{fig6(a)}, it can be observed that the BFS-AO method converges faster than the BFS-PDS method across various transmission power levels, and ultimately converges to similar performance.
	In Fig.~\subref*{fig6(b)}, we evaluate the convergence behavior of the MAPPO algorithm under two different initialization strategies. 
	Compared to the random initialization, the MAPPO algorithm utilizing the calculated initialization converges faster, which validates the effectiveness of initial observations provided by the BFS-AO method.
	It is also observed that networks converge more rapidly when the number of RIS elements is small. The reason is that, as the dimension of observation and action become larger, the network parameters increase and the multi-agent environment becomes more complex, which effects the convergence speed.
	
	\subsection{Generalizability}
	\begin{figure*}[htbp]
		\centering
		\subfloat[Generalizability to $P_{\max}$]{
			\includegraphics[width=0.45\textwidth]{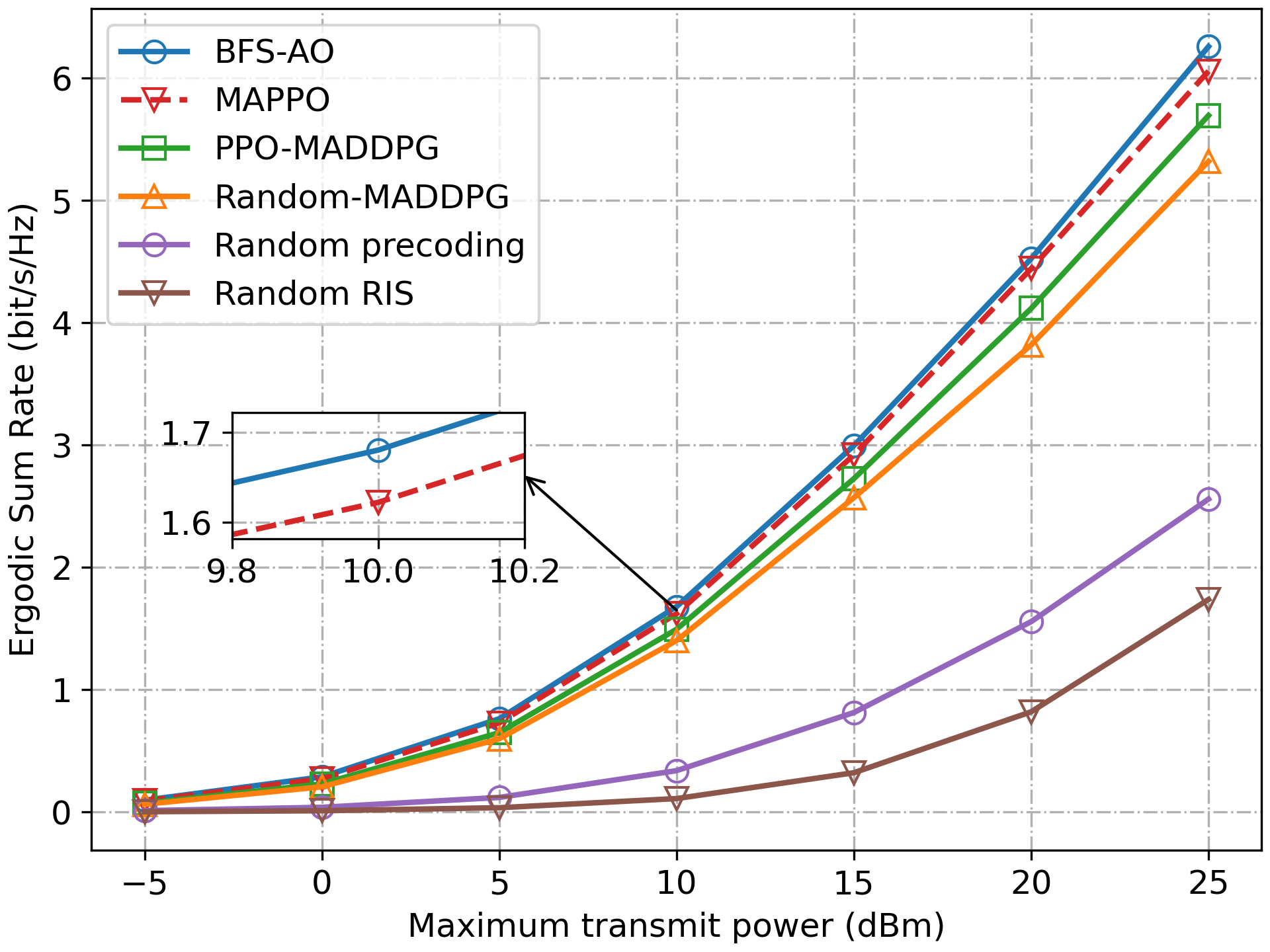}
			\label{fig7(a)}
			\vspace{-0.25cm}}
		\hfill
		\subfloat[Generalizability to $K_{k, l}$]{
			\includegraphics[width=0.45\textwidth]{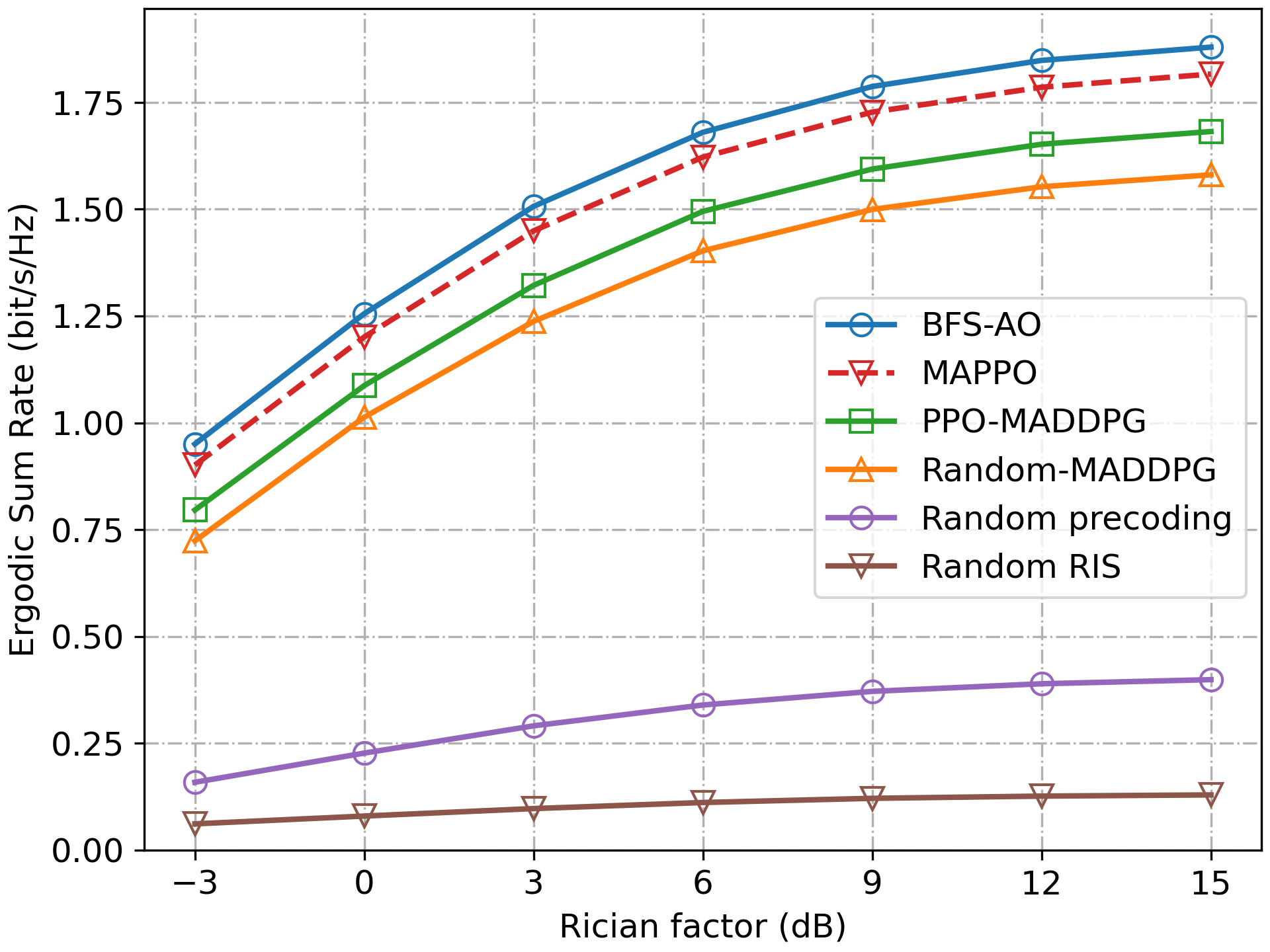}
			\label{fig7(b)}
			\vspace{-0.25cm}}
		\caption{Generalizability of the MAPPO algorithm in a distributed RIS-aided system.}
		\label{fig7}
	\end{figure*}

	Fig.~\ref{fig7} illustrates the generalizability of the MAPPO algorithm to the transmission power and Rician factor. In Fig.~\subref*{fig7(a)}, the MAPPO algorithm is trained at $P_{\max}=10$ dBm and tested in different scenarios with the maximum transmit power varying from $-5$ dBm to $25$ dBm.
	As can be observed, the MAPPO algorithm always approaches the BFS-AO method across the varying maximum transmit power, which verifies the strong generalizability of the MAPPO algorithm to $P_{\max}$. 	
	Compared to the PPO-MADDPG method, the effectiveness of MAPPO in joint precoding design is also validated, which hinges on two critical elements: the CTDE paradigm to overcome the non-stationarity of the environment, and the well-designed observation and action to ensure the algorithm's convergence.
	
	The impact of $K_{k, l}$ on the ergodic sum rate is shown in Fig.~\subref*{fig7(b)}. Given that the BS-RIS links usually remains steady due to the fixed locations, we set $K_l=6$ dB and adjust $K_{k,l}$ to evaluate the performance. The MAPPO algorithm trained at $K_{k,l}=6$ dB is tested in other scenarios with different $K_{k,l}$. It is seen that the performance of all algorithms exploiting statistical CSI improve as $K_{k,l}$ increases, and the MAPPO algorithm approaches the BFS-AO method with the varying Rician factor, which validates the generalizability of the MAPPO algorithm to $K_{k,l}$. 
	In addition, the comparison between the PPO-MADDPG and Random-MADDPG methods demonstrates the superiority of the scheduling agent over random scheduling.

	\subsection{User Scalability}	
	\begin{figure*}[htbp]
		\centering
		\subfloat[Scalability to $\hat{K}$]{
			\includegraphics[width=0.45\textwidth]{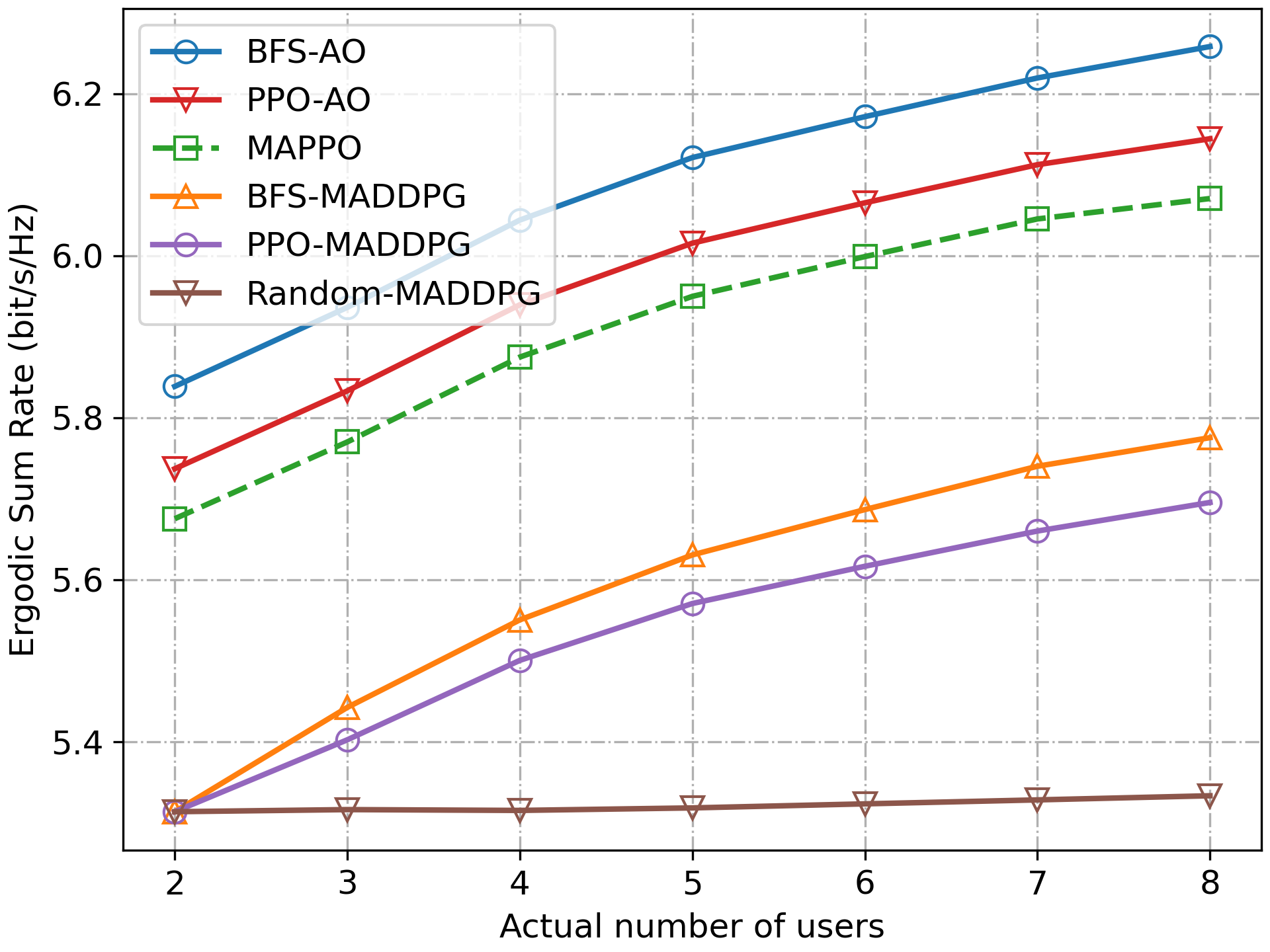}
			\label{fig8(a)}
			\vspace{-0.25cm}}
		\hfill
		\subfloat[Scalability to $K$]{
			\includegraphics[width=0.45\textwidth]{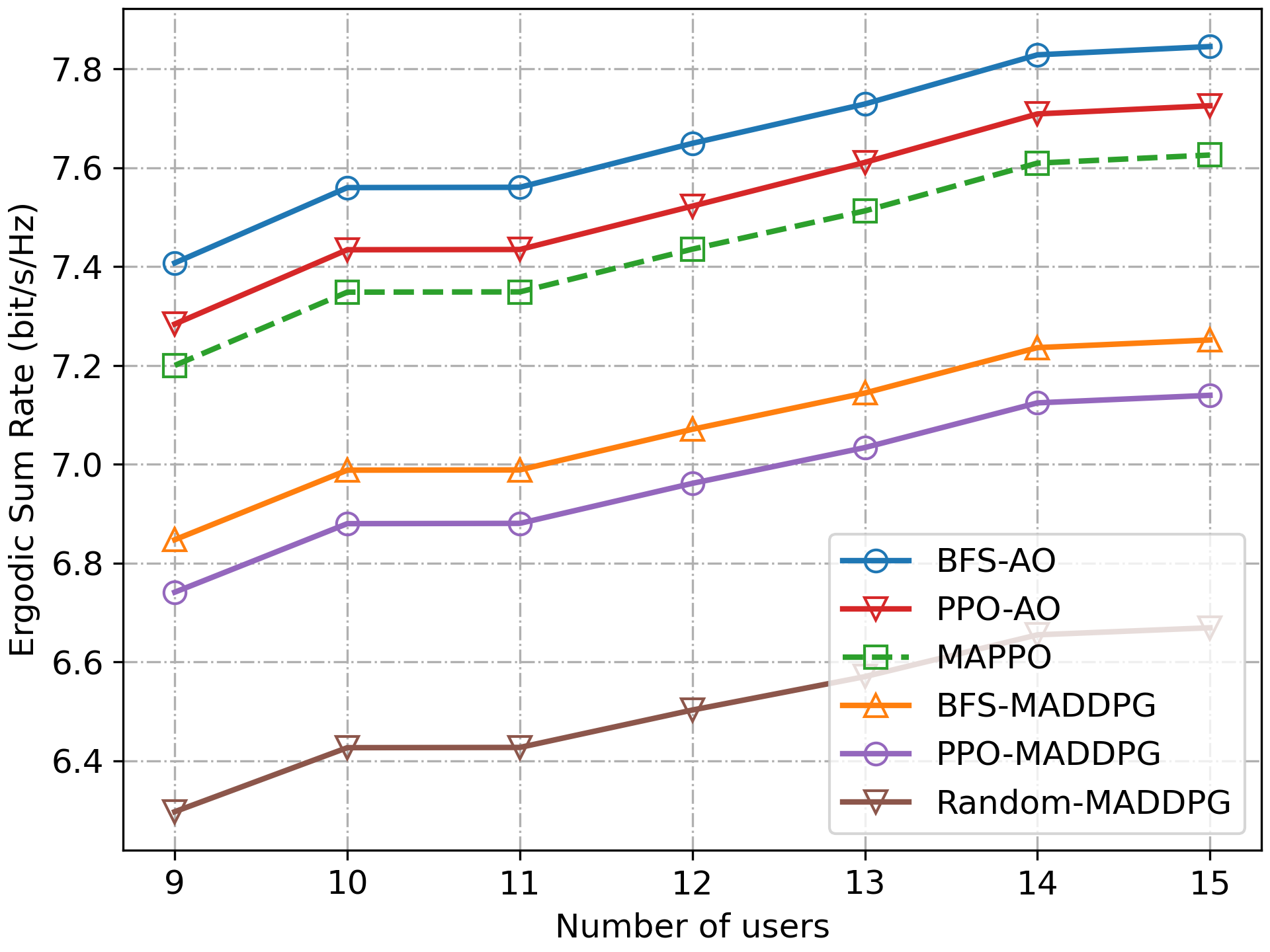}
			\label{fig8(b)}
			\vspace{-0.25cm}}
		\caption{User scalability of the MAPPO algorithm in a distributed RIS-aided system.}
		\label{fig8}
	\end{figure*}
	
	Fig.~\ref{fig8} shows the user scalability of the MAPPO algorithm to $\hat{K}$ and $K$. In Fig.~\subref*{fig8(a)}, we aim to show that the trained MAPPO algorithm can be applied directly to scenarios where the actual number of users varies. The case of $\hat{K}>K$ has been discussed in Section \ref{subsec2}, and next we simulate the scenario where $\hat{K} \leq K$ by adding virtual users with all-zero channel matrices. The MAPPO algorithm is trained at $P_{\max}=10$ dBm and $K=8$, and then applied under the condition of $P_{\max}=25$ dBm. It can be found that the performance of most algorithms declines as $\hat{K}$ decreases, while the Random-MADDPG method exploiting random scheduling is not much affected. 
	Furthermore, the performance of the MAPPO algorithm is comparable to the BFS-AO method with a gap no more than $0.2$ bit/s/Hz. A comparison of the BFS-MADDPG and PPO-MADDPG method reveals the effectiveness of the scheduling agent in user scheduling, and the performance gap between them becomes negligible when $\hat{K}$ decreases.  
	
	To verify the scalability to $K$, the MAPPO algorithm is trained at $P_{\max}=10$ dBm and $r=9$ m, and then applied under the condition of $P_{\max}=25$ dBm. With more added users, we retrain the algorithm in the same way. As shown in Fig.~\subref*{fig8(b)}, we can see that more users lead to better performance of all algorithms, and the MAPPO algorithm performs comparably to the PPO-AO method within a gap of $0.1$ bit/s/Hz.
	Moreover, the networks are retrained with different channels, which results in non-smooth performance curves of all algorithms.
	By comparing with benchmarks, the scalability of the MAPPO algorithm to $K$ is validated.
	
	\subsection{RIS Scalability}
	\begin{figure*}[htbp]
		\subfloat[Scalability to $N$]{
			\centering
			\includegraphics[width=0.45\textwidth]{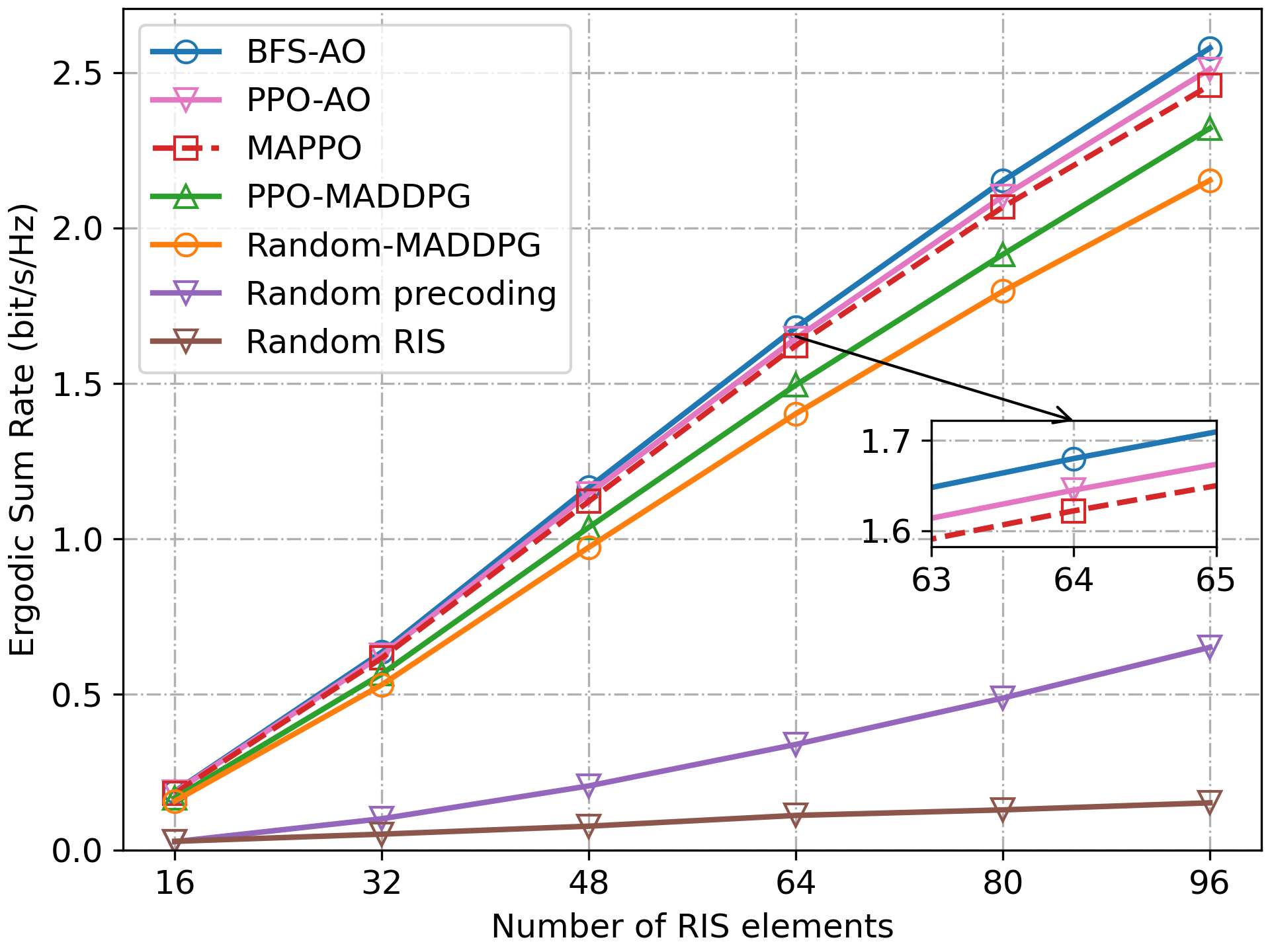}
			\label{fig9(a)}
			\vspace{-0.25cm}}
		\hfill
		\subfloat[Scalability to $L$]{
			\centering
			\includegraphics[width=0.45\textwidth]{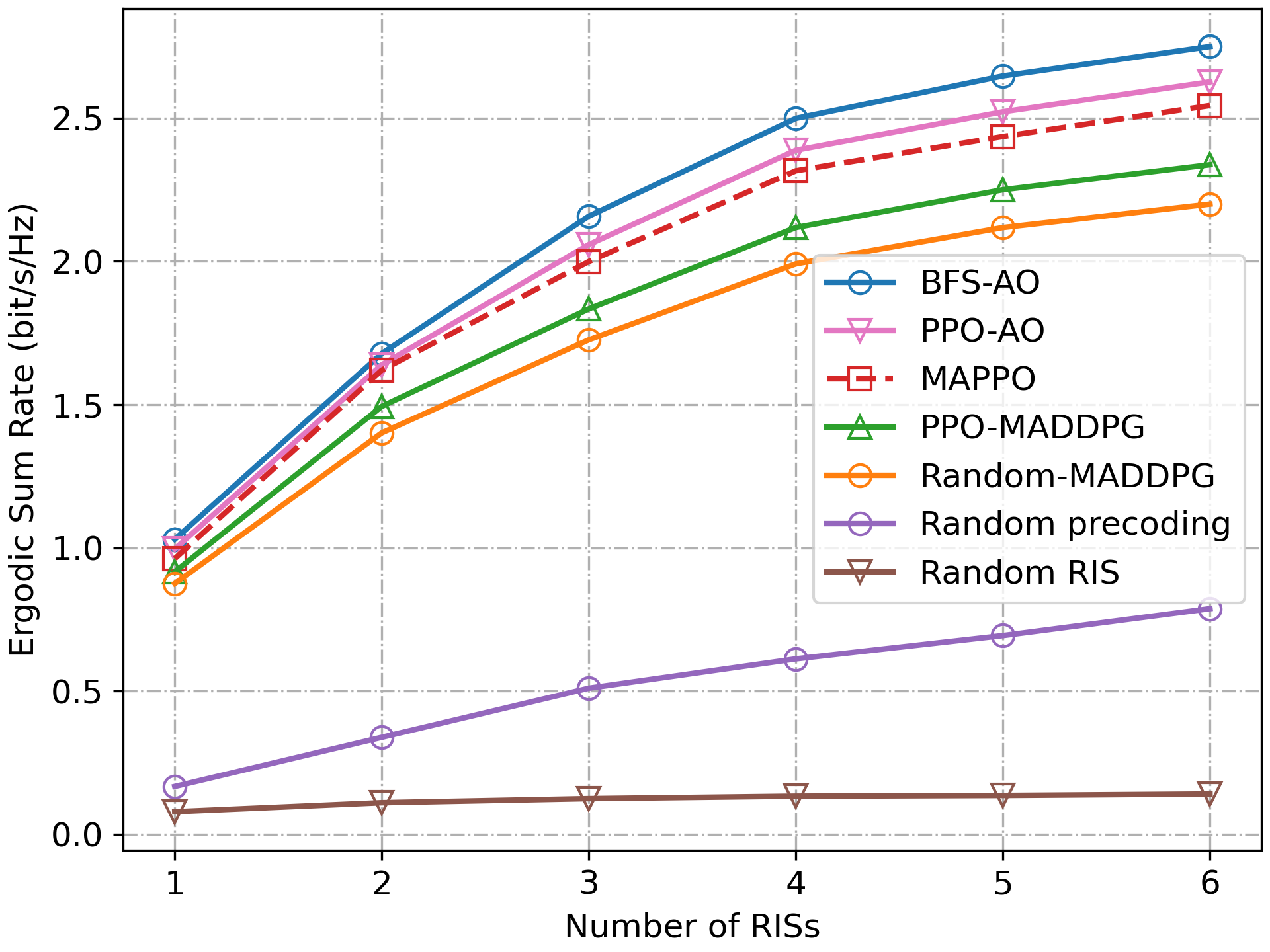}
			\label{fig9(b)}
			\vspace{-0.25cm}}
		\caption{RIS scalability of the MAPPO algorithm in a distributed RIS-aided system.}
		\label{fig9}
	\end{figure*}
	
	Fig.~\ref{fig9} shows the RIS scalability of the MAPPO algorithm to $N$ and $L$. In Fig.~\subref*{fig9(a)}, we evaluate the performance by varying the number of RIS elements, and the MAPPO algorithm is retrained to adapt different numbers of elements. 
	It can be easily found that within the range of $16$ and $96$, increasing $N$ leads to a improvement in the ergodic sum rate. Moreover, the BFS-AO method is closely approached by the MAPPO algorithm, which verifies the valid scalability to $N$. 
	In comparison with the ``Random precoding" and ``Random RIS" methods, we verify the significant effect of BS precoding and RIS precoding on enhancing the ergodic sum rate, respectively.
	
	To verify the scalability to $L$, four more RISs are located at ($90$ m, $30$ m, $10$ m) ($30$ m, $90$ m, $10$ m) ($70$ m, $100$ m, $10$ m) and ($100$ m, $70$ m, $10$ m), respectively. 
	In Fig.~\subref*{fig9(b)}, the MAPPO algorithm requires retraining as the number of RISs increases from one to six. It is worth noting that the MAPPO algorithm consistently maintains a similar growth trend to the BFS-AO method. 
	Thanks to the proposed observation design, 
	the observation dimension remains fixed as $N$ and $L$ increase, which enhances the scalability of the MAPPO algorithm.
	Furthermore, we observe a diminishing growth rate of the ergodic sum rate after the deployment of the fourth RIS, due to the increasing distance between subsequent RISs, the BS and users. 
	
	\subsection{User Fairness}
	\begin{figure}[htbp]
		\centering
		\subfloat[Ergodic sum rate]{
			\includegraphics[width=0.475\columnwidth]{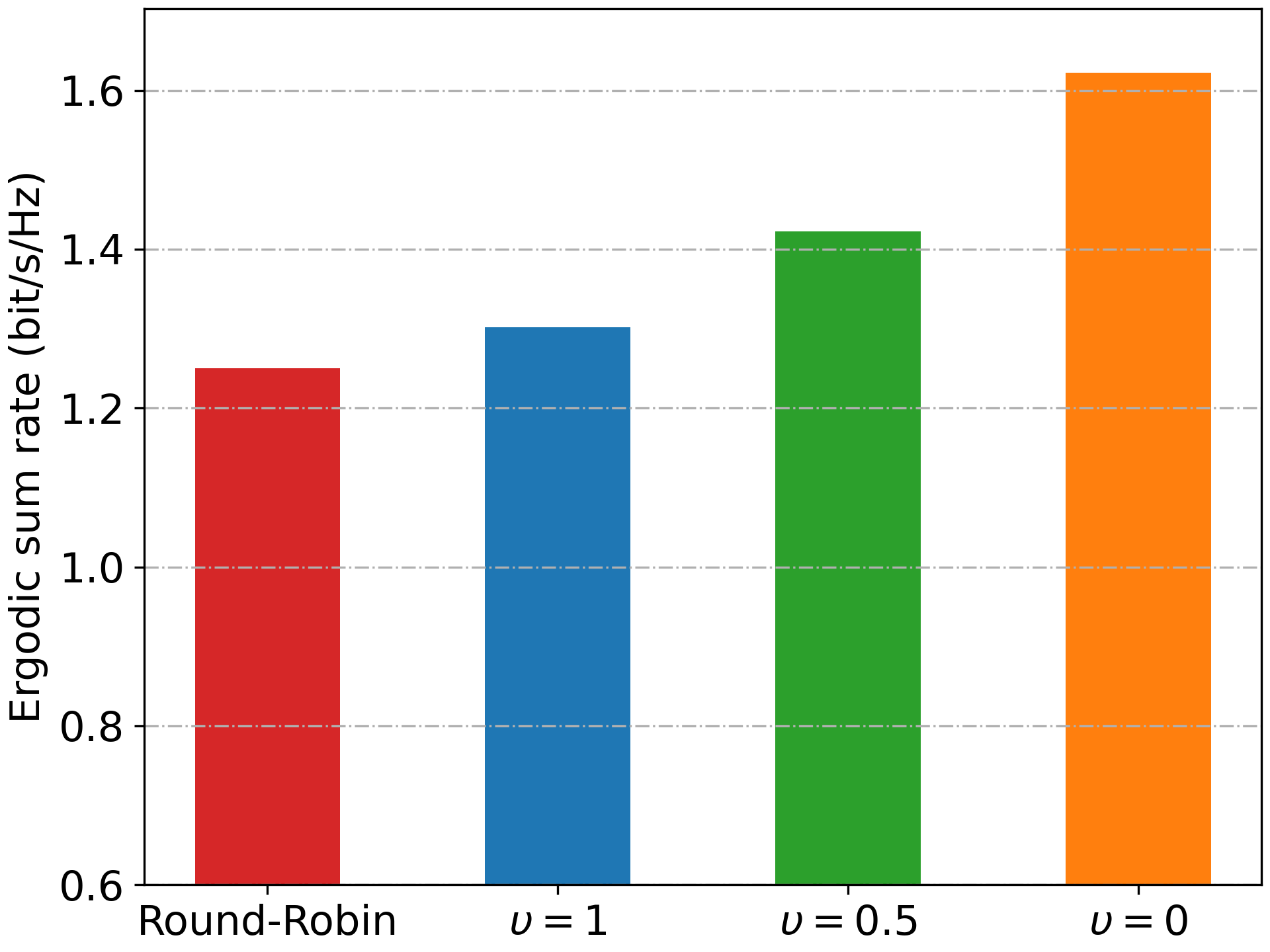}
			\label{fig10(a)}
			\vspace{-0.25cm}}
		\subfloat[JFI]{
			\includegraphics[width=0.475\columnwidth]{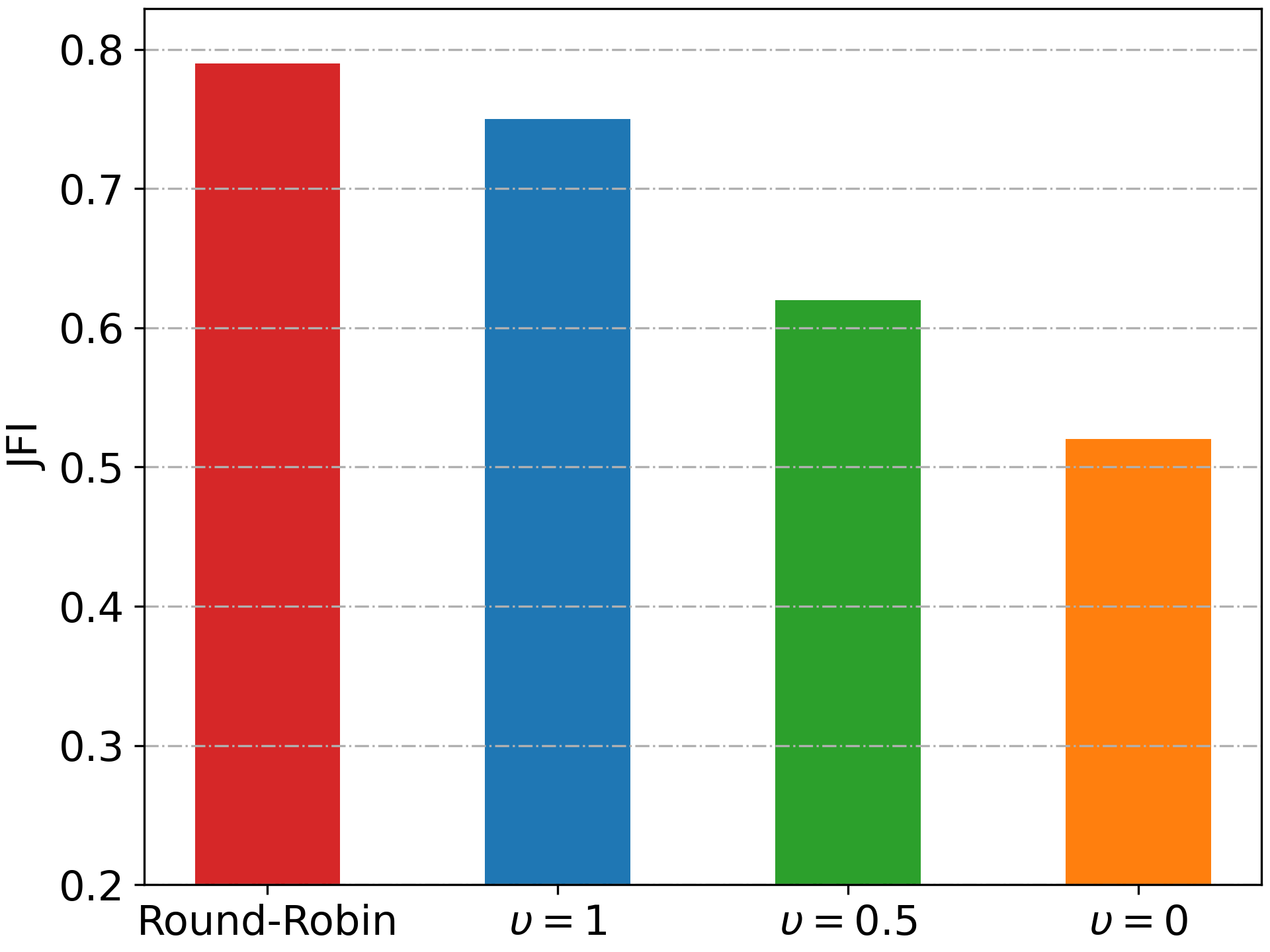}
			\label{fig10(b)}
			\vspace{-0.25cm}}
		\caption{Performance comparison of the ergodic sum rate and JFI.}
		\label{fig10}
	\end{figure}
	
	Fig.~\ref{fig10} compares the ergodic sum rate and JFI performance of the Round-Robin algorithm and MAPPO algorithm across varying $\upsilon$. 
	As can be observed, the MAPPO algorithm with $\upsilon=0$ achieves the highest ergodic sum rate but the lowest JFI. In comparison, the Round-Robin algorithm exhibits the highest JFI but the lowest ergodic sum rate. Moreover, the JFI of the MAPPO algorithm improves as $\upsilon$ increases, verifying that the proposed algorithm can favor fairness through adjusting the fairness weight factor.
	
	\subsection{Running Time}
	In the previous section, the MAPPO algorithm exhibits a computational complexity of $\mathcal{O}\bigl(\bigl(KM+MUL+NL+\binom{K}{U}\bigr)I + LI^2\bigr)$, which is much lower than that of the BFS-AO method $\mathcal{O}\bigl(I_1 \binom{K}{U}\bigl(I_2 K M^3 + I_3 K^2 L^2 N^2\bigr)\bigr)$, and the PPO-AO method $\mathcal{O}\bigl(\bigl(KM+\binom{K}{U}\bigr)I + I_1 \bigl(I_2 K M^3 + I_3 K^2 L^2 N^2\bigr) \bigr)$.
	In Table \ref{tab2}, we compare the running time of the above algorithms, while the performance ratios relative to the BFS-AO method are also given. 
	It can be seen that the MAPPO algorithm has significant advantages in the running time at the cost of slight performance loss, which also demonstrates the comparable performance of the distributed algorithm to centralized AO-based methods.
	By independently configuring RIS precoding, the distributed algorithm can eliminate the system signaling overhead of $BNL$, where $B$ denotes the number of bits for each transmitted symbol.
	With the growth of $N$, the increasing network dimensionality leads to a slight degradation in the performance ratio. 
	Furthermore, the running time of the BFS-AO method increases dramatically, while that of the MAPPO algorithm grows much more slowly as $N$ increases.
	\vspace{-0.1em}
	
	\begin{table}[htbp]
		\renewcommand{\arraystretch}{1.2}
		\setlength{\tabcolsep}{6pt}
		\caption{Performance And Running Time Comparison}
		\begin{center}
			\scalebox{1}{
			\begin{tabular}{|c|c|c|c|}
				\hline
				Algorithm&Parameter&Running time(ms)&Performance\\
				\hline
				\multirow{2}{*}{MAPPO}&$N=64$&1.23&96.43\%\\
				\cline{2-4} 
				&$N=96$&1.42&96.17\%\\
				\hline
				\multirow{2}{*}{PPO-AO}&$N=64$&113.48&97.91\%\\
				\cline{2-4} 
				&$N=96$&171.66&97.36\%\\
				\hline
				\multirow{2}{*}{BFS-AO}&$N=64$&1957.21&100.00\%\\
				\cline{2-4} 
				&$N=96$&2946.29&100.00\%\\
				\hline
			\end{tabular}}
			\label{tab2}
		\end{center}
	\end{table}
	
	\section{Conclusion}
	In this paper, we investigated the joint optimization of user scheduling and precoding for distributed RIS-aided systems. The closed-form approximation of the ergodic sum rate was firstly derived by exploiting statistical CSI. Then, the BFS-AO method was proposed and provided initial observations for the MADRL algorithm. 
	To reduce the computational complexity of the numerical optimization method, we proposed a data-driven scalable and generalizable MADRL framework. To be specific, the scheduling and BS precoding agents optimize user scheduling and active precoding, respectively, while passive precoding is determined by the corresponding RIS agent.
	After centralized training, we proposed a novel dynamic working process, which enables distributed RISs to configure their own passive precoding independently. 
 	Simulation results validated the strong generalizability of the MAPPO algorithm to environmental factors, including the BS transmit power, the Rician factor and the actual number of users, along with its excellent scalability to the number of users, RIS elements and RISs. In comparison with other benchmarks, the MAPPO algorithm shows superior performance with low computational complexity. 

\bibliographystyle{IEEEtran}
\bibliography{Reference}

\begin{thebibliography}{10}
\providecommand{\url}[1]{#1}
\csname url@samestyle\endcsname
\providecommand{\newblock}{\relax}
\providecommand{\bibinfo}[2]{#2}
\providecommand{\BIBentrySTDinterwordspacing}{\spaceskip=0pt\relax}
\providecommand{\BIBentryALTinterwordstretchfactor}{4}
\providecommand{\BIBentryALTinterwordspacing}{\spaceskip=\fontdimen2\font plus
\BIBentryALTinterwordstretchfactor\fontdimen3\font minus
  \fontdimen4\font\relax}
\providecommand{\BIBforeignlanguage}[2]{{%
\expandafter\ifx\csname l@#1\endcsname\relax
\typeout{** WARNING: IEEEtran.bst: No hyphenation pattern has been}%
\typeout{** loaded for the language `#1'. Using the pattern for}%
\typeout{** the default language instead.}%
\else
\language=\csname l@#1\endcsname
\fi
#2}}
\providecommand{\BIBdecl}{\relax}
\BIBdecl

\bibitem{wang2023multiagent}
Y.~Wang, X.~Li, N.~Gao, and S.~Jin, ``Multi-agent {DRL} based transmission
  optimization for distributed {RIS}-aided communication systems,'' in
  \emph{Proc. {IEEE} Int. Conf. Commun. Technol. (ICCT)}, Oct. 2023.

\bibitem{wu2020towards}
Q.~Wu and R.~Zhang, ``Towards smart and reconfigurable environment: Intelligent
  reflecting surface aided wireless network,'' \emph{IEEE Commun. Mag.},
  vol.~58, no.~1, pp. 106--112, Jan. 2020.

\bibitem{liu2022integrated}
F.~Liu \emph{et~al.}, ``Integrated sensing and communications: Toward
  dual-functional wireless networks for {6G} and beyond,'' \emph{IEEE J. Sel.
  Areas Commun.}, vol.~40, no.~6, pp. 1728--1767, Jun. 2022.

\bibitem{akyildiz2022terahertz}
I.~F. Akyildiz, C.~Han, Z.~Hu, S.~Nie, and J.~M. Jornet, ``Terahertz band
  communication: An old problem revisited and research directions for the next
  decade,'' \emph{IEEE Trans. Commun.}, vol.~70, no.~6, pp. 4250--4285, Jun.
  2022.

\bibitem{wang2022reconfigurable}
J.~Wang \emph{et~al.}, ``Reconfigurable intelligent surface: Power consumption
  modeling and practical measurement validation,'' \emph{arXiv},
  arXiv:2211.00323, 2022.

\bibitem{sang2022coverage}
J.~Sang \emph{et~al.}, ``Coverage enhancement by deploying {RIS} in {5G}
  commercial mobile networks: Field trials,'' \emph{IEEE Trans. Commun.},
  vol.~31, no.~1, pp. 172--180, Feb. 2024.

\bibitem{guo2020weighted}
H.~Guo, Y.-C. Liang, J.~Chen, and E.~G. Larsson, ``Weighted sum-rate
  maximization for reconfigurable intelligent surface aided wireless
  networks,'' \emph{IEEE Trans. Wireless Commun.}, vol.~19, no.~5, pp.
  3064--3076, May 2020.

\bibitem{wu2019intelligent}
Q.~Wu and R.~Zhang, ``Intelligent reflecting surface enhanced wireless network
  via joint active and passive beamforming,'' \emph{IEEE Trans. Wireless
  Commun.}, vol.~18, no.~11, pp. 5394--5409, Nov. 2019.

\bibitem{huang2018reconfigurable}
C.~Huang, A.~Zappone, G.~Alexandropoulos, M.~Debbah, and C.~Yuen,
  ``Reconfigurable intelligent surfaces for energy efficiency in wireless
  communication,'' \emph{IEEE Trans. Wireless Commun.}, vol.~18, no.~8, pp.
  4157--4170, Aug. 2019.

\bibitem{li2023risenhanced}
X.~Li, L.~Jiang, C.~Luo, Y.~Han, M.~Matthaiou, and S.~Jin, ``{RIS}-enhanced
  multi-cell downlink transmission using statistical channel state
  information,'' \emph{Sci. China Inf. Sci.}, vol.~66, no.~11, Sep. 2023.

\bibitem{zhang2022reconfigurable}
Y.~Zhang, J.~Zhang, M.~D. Renzo, H.~Xiao, and B.~Ai, ``Reconfigurable
  intelligent surfaces with outdated channel state information: Centralized vs.
  distributed deployments,'' \emph{IEEE Trans. Commun.}, vol.~70, no.~4, pp.
  2742--2756, Apr. 2022.

\bibitem{li2020weighted}
Z.~Li, M.~Hua, Q.~Wang, and Q.~Song, ``Weighted sum-rate maximization for
  multi-{IRS} aided cooperative transmission,'' \emph{IEEE Wireless Commun.
  Lett.}, vol.~9, no.~10, pp. 1620--1624, Oct. 2020.

\bibitem{abrardo2021intelligent}
A.~Abrardo, D.~Dardari, and M.~D. Renzo, ``Intelligent reflecting surfaces:
  Sum-rate optimization based on statistical position information,'' \emph{IEEE
  Trans. Commun.}, vol.~69, no.~10, pp. 7121--7136, Oct. 2021.

\bibitem{gao2021distributed}
Y.~Gao, J.~Xu, W.~Xu, D.~W.~K. Ng, and M.-S. Alouini, ``Distributed {IRS} with
  statistical passive beamforming for {MISO} communications,'' \emph{IEEE
  Wireless Commun. Lett.}, vol.~10, no.~2, pp. 221--225, Feb. 2021.

\bibitem{nadeem2021opportunistic}
Q.-U.-A. Nadeem, A.~Chaaban, and M.~Debbah, ``Opportunistic beamforming using
  an intelligent reflecting surface without instantaneous {CSI},'' \emph{IEEE
  Wireless Commun. Lett.}, vol.~10, no.~1, pp. 146--150, Jan. 2021.

\bibitem{jiang2023joint}
L.~Jiang, X.~Li, M.~Matthaiou, and S.~Jin, ``Joint user scheduling and phase
  shift design for {RIS} assisted multi-cell {MISO} systems,'' \emph{IEEE
  Wireless Commun. Lett.}, vol.~12, no.~3, pp. 431--435, Mar. 2023.

\bibitem{mei2021performance}
W.~Mei and R.~Zhang, ``Performance analysis and user association optimization
  for wireless network aided by multiple intelligent reflecting surfaces,''
  \emph{IEEE Trans. Commun.}, vol.~69, no.~9, pp. 6296--6312, Sep. 2021.

\bibitem{liu2023joint}
S.~Liu, R.~Liu, M.~Li, Y.~Liu, and Q.~Liu, ``Joint {BS}-{RIS}-user association
  and beamforming design for {RIS}-assisted cellular networks,'' \emph{IEEE
  Trans. Veh. Technol.}, vol.~72, no.~5, pp. 6113--6128, May 2023.

\bibitem{amiriara2022irsuser}
H.~Amiriara, F.~Ashtiani, M.~Mirmohseni, and M.~Nasiri-Kenari, ``{IRS}-user
  association in {IRS}-aided {MISO} wireless networks: Convex optimization and
  machine learning approaches,'' \emph{IEEE Trans. Veh. Technol.}, vol.~72,
  no.~11, pp. 14\,305 -- 14\,316, Nov. 2023.

\bibitem{Al2022Reconfigurable}
A.~Al-Hilo, M.~Samir, M.~Elhattab, C.~Assi, and S.~Sharafeddine,
  ``Reconfigurable intelligent surface enabled vehicular communication: Joint
  user scheduling and passive beamforming,'' \emph{IEEE Trans. Veh. Technol.},
  vol.~71, no.~3, pp. 2333--2345, Mar. 2022.

\bibitem{taha2021enabling}
A.~Taha, M.~Alrabeiah, and A.~Alkhateeb, ``Enabling large intelligent surfaces
  with compressive sensing and deep learning,'' \emph{IEEE Access}, vol.~9, pp.
  44\,304--44\,321, Mar. 2021.

\bibitem{hu2021reconfigurable}
X.~Hu, C.~Masouros, and K.-K. Wong, ``Reconfigurable intelligent surface aided
  mobile edge computing: From optimization-based to location-only
  learning-based solutions,'' \emph{IEEE Trans. Commun.}, vol.~69, no.~6, pp.
  3709--3725, Mar. 2021.

\bibitem{song2021unsupervised}
H.~Song, M.~Zhang, J.~Gao, and C.~Zhong, ``Unsupervised learning-based joint
  active and passive beamforming design for reconfigurable intelligent surfaces
  aided wireless networks,'' \emph{IEEE Commun. Lett.}, vol.~25, no.~3, pp.
  892--896, Mar. 2021.

\bibitem{jin2024lowcomplexity}
W.~Jin, J.~Zhang, C.-K. Wen, S.~Jin, X.~Li, and S.~Han, ``Low-complexity joint
  beamforming for {RIS}-assisted {MU}-{MISO} systems based on model-driven deep
  learning,'' \emph{IEEE Trans. Wireless Commun.}, vol.~23, no.~7, pp.
  6968--6982, Jul. 2024.

\bibitem{feng2020deep}
K.~Feng, Q.~Wang, X.~Li, and C.-K. Wen, ``Deep reinforcement learning based
  intelligent reflecting surface optimization for {MISO} communication
  systems,'' \emph{IEEE Wireless Commun. Lett.}, vol.~9, no.~5, pp. 745--749,
  May 2020.

\bibitem{zhang2023a}
H.~Zhang, X.~Li, N.~Gao, X.~Yi, and S.~Jin, ``A deep reinforcement learning
  approach to two-timescale transmission for {RIS}-aided multiuser {MISO}
  systems,'' \emph{IEEE Wireless Commun. Lett.}, vol.~12, no.~8, pp.
  1444--1448, Aug. 2023.

\bibitem{wang2021hybrid}
Q.~Wang, X.~Li, S.~Jin, and Y.~Chen, ``Hybrid beamforming for {mmWave}
  {MU}-{MISO} systems exploiting multi-agent deep reinforcement learning,''
  \emph{IEEE Wireless Commun. Lett.}, vol.~10, no.~5, pp. 1046--1050, May 2021.

\bibitem{naderializadeh2021resource}
N.~Naderializadeh, J.~J. Sydir, M.~Simsek, and H.~Nikopour, ``Resource
  management in wireless networks via multi-agent deep reinforcement
  learning,'' \emph{IEEE Trans. Wireless Commun.}, vol.~20, no.~6, pp.
  3507--3523, Jun. 2021.

\bibitem{li2016statistical}
X.~Li, S.~Jin, H.~A. Suraweera, J.~Hou, and X.~Gao, ``Statistical 3-{D}
  beamforming for large-scale {MIMO} downlink systems over {Rician} fading
  channels,'' \emph{IEEE Trans. Commun.}, vol.~64, no.~4, pp. 1529--1543, Apr.
  2016.

\bibitem{luo2021reconfigurable}
C.~Luo, X.~Li, S.~Jin, and Y.~Chen, ``Reconfigurable intelligent
  surface-assisted multi-cell {MISO} communication systems exploiting
  statistical {CSI},'' \emph{IEEE Wireless Commun. Lett.}, vol.~10, no.~10, pp.
  2313--2317, Oct. 2021.

\bibitem{tampuu2017multiagent}
A.~Tampuu \emph{et~al.}, ``Multiagent cooperation and competition with deep
  reinforcement learning,'' \emph{PloS One}, vol.~12, no.~4, Apr. 2017.

\bibitem{witt2020is}
C.~S. Witt \emph{et~al.}, ``Is independent learning all you need in the
  {StarCraft} multi-agent challenge?'' \emph{arXiv}, arXiv:2011.09533, 2020.

\bibitem{sukhbaatar2016learning}
S.~Sukhbaatar, A.~Szlam, and R.~Fergus, ``Learning multiagent communication
  with backpropagation,'' in \emph{Proc. Neural Inf. Proces. Syst. (NeurIPS)},
  Dec. 2016, pp. 2244--2252.

\bibitem{jiang2018learning}
J.~Jiang and Z.~Lu, ``Learning attentional communication for multi-agent
  cooperation,'' in \emph{Proc. Neural Inf. Proces. Syst. (NeurIPS)}, Dec.
  2018, pp. 7265--7275.

\bibitem{yu2022the}
C.~Yu \emph{et~al.}, ``The surprising effectiveness of {PPO} in cooperative
  multi-agent games,'' in \emph{Proc. Neural Inf. Proces. Syst. (NeurIPS)},
  Dec. 2022, pp. 24\,611--24\,624.

\bibitem{foerster2018counterfactual}
J.~N. Foerster, G.~Farquhar, T.~Afouras, N.~Nardelli, and S.~Whiteson,
  ``Counterfactual multi-agent policy gradients.'' in \emph{Proc. AAAI Conf.
  Artif. Intell.}, Feb. 2018, pp. 2974--2982.

\bibitem{sunehag2017valuedecomposition}
P.~Sunehag \emph{et~al.}, ``Value-decomposition networks for cooperative
  multi-agent learning,'' in \emph{Proc. Adaptive Agents Multi-Agent Syst.
  (AAMAS)}, May 2017.

\bibitem{rashid2020monotonic}
T.~Rashid, M.~Samvelyan, C.~S.~D. Witt, G.~Farquhar, J.~Foerster, and
  S.~Whiteson, ``Monotonic value function factorisation for deep multi-agent
  reinforcement learning,'' \emph{J. Mach. Learn. Res.}, vol.~21, no. 178, pp.
  1--51, 2020.

\bibitem{christianos2021scaling}
F.~Christianos, G.~Papoudakis, A.~Rahman, and S.~V. Albrecht, ``Scaling
  multi-agent reinforcement learning with selective parameter sharing,'' in
  \emph{Proc. Int. Conf. Mach. Learn. (ICML)}, Jul. 2021, pp. 1989--1998.

\bibitem{schulman2017proximal}
J.~Schulman, F.~Wolski, P.~Dhariwal, A.~Radford, and O.~Klimov, ``Proximal
  policy optimization algorithms,'' \emph{arXiv}, arXiv:1707.06347, 2017.

\bibitem{lin2021tensorbased}
Y.~Lin, S.~Jin, M.~Matthaiou, and X.~You, ``Tensor-based algebraic channel
  estimation for hybrid {IRS}-assisted {MIMO}-{OFDM},'' \emph{IEEE Trans.
  Wireless Commun.}, vol.~20, no.~6, pp. 3770--3784, Jun. 2021.

\bibitem{lin2022channel}
------, ``Channel estimation and user localization for {IRS}-assisted
  {MIMO}-{OFDM} systems,'' \emph{IEEE Trans. Wireless Commun.}, vol.~21, no.~4,
  pp. 2320--2335, Apr. 2022.

\bibitem{li2018user}
X.~Li, T.~Sun, N.~Qin, S.~Jin, and X.~Gao, ``User scheduling for downlink
  {FD}-{MIMO} systems under {Rician} fading exploiting statistical {CSI},''
  \emph{Sci. China Inf. Sci.}, vol.~61, no.~8, Apr. 2018.

\bibitem{sediq2013optimal}
A.~B. Sediq, R.~H. Gohary, R.~Schoenen, and H.~Yanikomeroglu, ``Optimal
  tradeoff between sum-rate efficiency and {Jain's} fairness index in resource
  allocation,'' \emph{IEEE Trans. Wireless Commun.}, vol.~12, no.~7, pp.
  3496--3509, Jul. 2013.

\bibitem{zhang2021joint}
Z.~Zhang and L.~Dai, ``A joint precoding framework for wideband reconfigurable
  intelligent surface-aided cell-free network,'' \emph{IEEE Trans. Signal
  Process.}, vol.~69, pp. 4085--4101, Jun. 2021.

\bibitem{li2024optimal}
X.~Li and S.~Bi, ``Optimal {AI} model splitting and resource allocation for
  device-edge co-inference in multi-user wireless sensing systems,'' \emph{IEEE
  Trans. Wireless Commun.}, vol.~23, no.~9, pp. 11\,094--11\,108, Sep. 2024.

\end{thebibliography}
\end{document}